\documentclass[a4paper,11pt]{article}
\pdfoutput=1
\usepackage{jcappub}
\usepackage{graphicx}
\usepackage{amsmath}
\usepackage{stmaryrd}
\usepackage[dvicolors]{xcolor}
\usepackage{hyperref}
\usepackage[normalem]{ulem}
\begin{document}
\title{Prospects for realtime characterization of core-collapse supernova and neutrino properties}
\author[a,d,1]{Meriem Bendahman,\note{These authors contributed equally to this work.}}
\author[a,1]{Isabel Goos,}
\author[a]{Joao Coelho,}
\author[b,c]{Matteo Bugli,}
\author[a]{Alexis Coleiro,}
\author[a,2]{Sonia El Hedri,\note{Corresponding author}}
\author[c]{Thierry Foglizzo,}
\author[a]{Davide Franco,} 
\author[c]{J\'er\^ome Guilet,}
\author[a]{Antoine Kouchner,}
\author[f]{Rapha\"el Raynaud,}
\author[d,e]{and Yahya Tayalati}
\affiliation[a]{Université Paris Cité, CNRS, Astroparticule et Cosmologie, F-75013 Paris, France}
\affiliation[b]{Dipartimento di Fisica, Università di Torino, I-10125 Torino, Italy}
\affiliation[c]{Universit\'e Paris-Saclay, Universit\'e Paris Cit\'e, CEA, CNRS, AIM,
91191, Gif-sur-Yvette, France}
\affiliation[d]{Faculty of Sciences, Mohammed V University in Rabat, Avenue des Nations Unies, Rabat, Morocco}
\affiliation[e]{Institute of Applied Physics, Mohammed VI Polytechnic University, Ben Guerir, Morocco}
\affiliation[f]{Université Paris Cité, Université Paris-Saclay, CEA, CNRS, AIM, F-91191 Gif-sur-Yvette, France}
\emailAdd{bendahman@apc.in2p3.fr}
\emailAdd{goos@apc.in2p3.fr}
\emailAdd{jcoelho@apc.in2p3.fr} 
\emailAdd{matteo.bugli@unito.it}
\emailAdd{coleiro@apc.in2p3.fr}
\emailAdd{elhedri@apc.in2p3.fr} 
\emailAdd{foglizzo@cea.fr}
\emailAdd{davide.franco@apc.in2p3.fr}
\emailAdd{jerome.guilet@cea.fr}
\emailAdd{kouchner@apc.in2p3.fr}
\emailAdd{raphael.raynaud@cea.fr}
\emailAdd{yahya.tayalati@cern.ch}
\abstract{Core-collapse supernovae (CCSNe) offer extremely valuable insights into the dynamics of galaxies. Neutrino time profiles from CCSNe, in particular, could reveal unique details about collapsing stars and particle behavior in dense environments. However, CCSNe in our galaxy and the Large Magellanic Cloud are rare and only one supernova neutrino observation has been made so far. To maximize the information obtained from the next Galactic CCSN, it is essential to combine analyses from multiple neutrino experiments in real time and transmit any relevant information to electromagnetic facilities within minutes. Locating the CCSN, in particular, is challenging, requiring disentangling CCSN localization information from observational features associated with the properties of the supernova progenitor and the physics of the neutrinos. Yet, being able to estimate the progenitor distance from the neutrino signal would be of great help for the optimisation of the electromagnetic follow-up campaign that will start soon after the propagation of the neutrino alert. Existing CCSN distance measurement algorithms based on neutrino observations hence rely on the assumption that neutrino properties can be described by the Standard Model. This paper presents a swift and robust approach to extract CCSN and neutrino physics information, leveraging diverse next-generation neutrino detectors to counteract potential measurement biases from Beyond the Standard Model effects.}
\maketitle
\flushbottom
\section{Introduction}
Core-collapse supernovae (CCSNe) play a pivotal role in shaping the dynamics and composition of galaxies \cite{martizzi2015,dubois2015,hopkins2018}. These cataclysmic events not only result in spectacular explosions but also give rise to essential cosmic phenomena such as the production of heavy elements through explosive nucleosynthesis and the formation of neutron stars and stellar black holes. These compact objects, in turn, may serve as the seeds for extremely powerful cosmic phenomena like Active Galactic Nuclei and Gamma-Ray Bursts. Understanding the underlying mechanism of CCSN explosions and the link between their outcomes and the properties of their progenitors is thus of paramount importance.

Despite their profound implications, CCSNe are not fully understood yet~\cite{janka2012,janka2017,muller2020}, as their onset takes place deep inside the star's core,  a region concealed beneath opaque layers, where matter is subjected to extreme conditions. In addition to hosting the key processes triggering the CCSN, this core region provides a unique laboratory for probing high-density, high-magnetic-field, high-temperature environments and the possibility of exploring physics beyond the Standard Model (SM). However, the explosion mechanism is also extremely difficult to model. Although numerical simulations have made substantial progress in modeling the core of massive stars, accurately replicating the conditions required for a successful CCSN explosion remains a formidable computational challenge. Consequently, there is an imperative need for direct observational insights.

During a CCSN, the star's core emits an intense burst of neutrinos over a short but critical timespan. These neutrinos are generated through nuclear reactions triggered by the collapse and the shockwave's formation and propagation. Due to their feeble interactions with matter, neutrinos preserve most of their initial information during their journey to Earth, offering a unique window into various phases of the CCSN, with millisecond-level precision. Moreover, neutrinos are expected to exit the star minutes to hours before the CCSN becomes visible electromagnetically. The detection of the $\mathcal{O}(10~\mathrm{MeV})$ thermal neutrinos emitted by a CCSN would hence considerably narrow the search time window for gravitational wave detectors and provide an early warning for optical telescopes. However, the detection of CCSN neutrinos is challenging and limited to nearby supernovae in the Milky Way or the Large Magellanic Cloud. Not only these supernovae are extremely rare, with about two expected per century~\cite{snrate_galaxy}, but they can also elude optical telescopes if they are located behind the Galactic Center~\cite{distance_dust}. If a CCSN occurs, it is therefore imperative to maximize the amount of information collected in real time by combining observations from all sensitive neutrino experiments. The Supernova Early Warning System (SNEWS) serves this purpose~\cite{snews,snews2}. This alert system currently receives data from ten neutrino experiments, extracts information about the CCSN progenitor, and sends it to telescopes. It notably deploys algorithms to determine the CCSN's angular position (and the associated error box) and its distance from Earth within minutes after the reception of neutrino data. The CCSN’s angular position and distance estimates based on neutrino observations, if estimated quickly enough, will play a major role in optimizing the electromagnetic follow-up campaigns that will start immediately after the neutrino alert, in order to locate and study the supernova's electromagnetic counterpart. In particular, depending on the size of the error box, the search for the counterpart will be performed either using large field-of-view / lower sensitivity instruments or smaller field-of-view facilities which are generally more sensitive (see e.g.~\cite{distance_dust}). Similarly, estimating the distance to the source will constrain whether the supernova exploded in the foreground or behind the Galactic Center. Such information is fundamental in favouring infrared rather than optical observations, due to the extinction in the Galactic plane~\cite{distance_dust,beacom}.

A critical challenge for CCSN alert systems, like SNEWS, is disentangling within minutes three categories of information imprinted in neutrino observations: CCSN localization (angular position and distance), CCSN properties (e.g., neutron star Equation of State, progenitor mass, density profile), and neutrino properties (e.g., mass ordering, interaction nature). 
While only the first is time-sensitive, neglecting new physics phenomena such as neutrino two-body decays, if they are present, or making wrong assumptions about the ordering of neutrino masses could affect the estimations of the properties and the location of the CCSN progenitor.

In particular, the detected neutrino signal's intensity, crucial for CCSN distance measurements, depends significantly on both CCSN and neutrino properties. The degeneracy between CCSN distance and CCSN progenitor type has been extensively studied~\cite{kachelriess,oconnor}, revealing that the flux of neutrinos during the first tens of milliseconds of a CCSN exhibits weak dependence on the CCSN progenitor's properties. These studies demonstrated that the electron neutrino emission peak during the supernova neutronization phase, often referred to as a ``standard candle'', offers the potential for CCSN distance measurements with sub-$10$\% precision. However, within the time frames considered in distance measurement algorithms, the neutrino fluxes highly depend on the neutrino's intrinsic properties, primarily their mass ordering (MO) in the Standard Model~\cite{adiabaticMSW}. Additionally, several new physics scenarios can significantly distort CCSN neutrino fluxes while still evading constraints from beam, solar, and atmospheric neutrino measurements~\cite{self-interactions,sterile,spinflip,ando_decay,degouvea}. Since current distance measurement techniques~\cite{kachelriess,segerlund} always assume the SM, the existence of new physics mechanisms could lead to significant biases, as will be shown in this work. While methods to constrain neutrino properties have been proposed for several Beyond the Standard Model (BSM) scenarios, these methods either consider specific CCSN progenitors~\cite{spinflip}, assume that the CCSN distance is known~\cite{sterile2}, or use neutrino energy measurements~\cite{degouvea,bayesian_espectrum} which are not possible with every detector. One generic characteristic signature of these new physics phenomena, however, is that their effect on neutrino observations depends on the neutrino type and flavor. The next generation of neutrino and dark matter experiments will be particularly suited to exploit this feature.

In the next decade, the landscape of neutrino experiments will be radically transformed: large-scale Water Cherenkov (WC) detectors such as Hyper-Kamiokande (HK), IceCube, and KM3NeT, sensitive to electron antineutrinos, will be complemented by the DUNE experiment, sensitive to electron neutrinos, and by large-scale detectors sensitive to Coherent Elastic Neutrino-Nucleus scattering (CE$\nu$NS) such as DarkSide-20k and, later, ARGO. Once completed, this panel of experiments will be sensitive to three diﬀerent ﬂavor combinations of supernova neutrinos. Combining these different detector types will provide a unique sensitivity to flavor-dependent phenomena over a wide range of supernova models and distances.

In this paper, we present a methodology to combine data from diverse neutrino experiments and concurrently extract information regarding CCSN progenitors, CCSN locations, and neutrino properties, relying solely on measured neutrino rates. We demonstrate this methodology using a new physics scenario involving neutrino two-body decays~\cite{degouvea} and illustrate how new physics mechanisms can bias CCSN distance measurements if only SM neutrino interactions are assumed. We also propose techniques to identify deviations from the Standard Model and characterize new physics scenarios, capitalizing on the complementarity between different types of neutrino experiments. These algorithms offer realtime analysis capabilities, making them suitable for alert systems and adaptable to a range of representative new physics models.

This article proceeds as follows. In Section~\ref{sec:methodology}, we discuss the neutrino experiments, CCSN progenitor models, and neutrino flavor conversion processes considered in our analyses. For the latter, we present both the current SM flavor-conversion scenario and an overview of possible new physics processes susceptible to affect the flavor composition of CCSN neutrinos, with an emphasis on neutrino two-body decays, which will be used as a case study of the impact of BSM interactions throughout this work. In Section~\ref{sec:likelihood_approach}, we propose a likelihood-based approach, inspired by previous studies of the time-dependence of CCSN neutrino luminosity spectra, to simultaneously constrain CCSN and neutrino parameters. We apply this approach to the determination of the neutrino MO and to CCSN distance measurements in Sections~\ref{sec:nmo_full} and \ref{sec:distance}, respectively. Then, Section~\ref{sec:decay_full} presents a model-dependent and a model-independent approach to identify and characterize the effects of physics beyond the SM on neutrino observations. We conclude in Section~\ref{sec:discussion}.

\section{Methodology}
\label{sec:methodology}
To evaluate the impact of the CCSN progenitor characteristics, the CCSN localization, and the progenitor properties on observations, we simulate the neutrino rates expected at large-scale next-generation neutrino experiments for a wide range of CCSN models. Our simulations incorporate the dominant neutrino flavor conversion mechanisms expected in the SM as well as an example of a new physics scenario that may affect the flux and flavor composition of the neutrinos arriving on Earth. In what follows, we describe our choice of models and experiments.

\subsection{CCSN models}
\label{subsec:models}
This analysis focuses on a comprehensive set of 149 progenitor models, originally devised and introduced by Segerlund \emph{et al}~\cite{model}. These models are the product of one-dimensional simulations modelling the evolution of isolated massive stars with solar metallicity. Their zero-age main sequence (ZAMS) masses span a broad spectrum, ranging from $9$ to $120$ solar masses. The neutrino signal was computed by Segerlund \emph{et al}~\cite{segerlund} with 1D core-collapse simulations using an energy-dependent neutrino transport scheme, state-of-the-art microphysical electron capture rates and neutrino-matter interactions. 
These models were designed to encompass a wide range of pre-supernova structures typically found in CCSNe with iron cores. In Segerlund \emph{et al}~\cite{segerlund}, they were used to identify observables whose values have a weak dependence on the CCSN model or can be easily parameterized.
Consequently, they constitute an ideal dataset for our analysis.

It is worth noting that the CCSN progenitors represented in these models exhibit highly mass-dependent occurrence rates in the Universe, with lighter stars significantly outnumbering their heavier counterparts. To account for this mass dependence in our work, we attribute prior probabilities to the different CCSN models. These prior probabilities are taken to be proportional to the Salpeter Initial Mass Function (IMF), which characterizes the distribution of stars according to their ZAMS masses~\cite{salpeter}:
\begin{align}
    p(M) \propto M^{-2.35}.
\end{align}
\subsection{Neutrino experiments}
\label{subsec:experiments}
For this analysis, we consider a set of large-scale experiments sensitive to CCSN neutrinos which are expected to be fully built and to start collecting data within the next decades. In order to lift the degeneracies between CCSN localization, CCSN properties, and neutrino properties, we consider not only WC detectors, which have been prominent actors in the search for CCSNe since the observation of SN1987A, but also experiments sensitive to different neutrino flavors. Since we want to emulate detector performances as realistically as possible, we consider only experiments whose detection efficiencies have been made public, either in the SnowGLoBES software~\cite{snowglobes} or in publications by the respective collaborations~\cite{darkside_sn}. The experiments we have selected for this analysis are projected to be sensitive to three distinct neutrino flavor combinations: $\bar{\nu}_e$, $\nu_e$, and the sum of all neutrino flavors. The detectors and their properties are summarized in table~\ref{tab:experiments}.
  \begin{table*}[h!]
  \begin{center}
        \begin{tabular}{ccccc}
            \hline
            \textbf{Experiment} & \parbox[t]{2.5cm}{\textbf{Detected}\\ \textbf{$\nu$ flavor}} & \parbox[t]{2.5cm}{\textbf{Total mass} \\\textbf{(kT)}} & \parbox[t]{2.5cm}{\textbf{Efficiency at} \\\textbf{$\mathbf{20}$~MeV~(\%)}} & \parbox[t]{2.5cm}{\textbf{Background}\\ \textbf{rate (Hz)}}\\
            \hline
            Hyper-Kamiokande & $\bar{\nu}_e$ & 260 & 100 & 0\\
            IceCube & $\bar{\nu}_e$ & $10^6$ & 4.8 & $1.5\times 10^6$\\
            KM3NeT & $\bar{\nu}_e$ & $2.1\times 10^5$ & 0.07 & $4.5\times 10^6$\\
            DUNE & $\nu_e$ & 40 & 100 & 0\\
            DarkSide-20k & all & 0.05 & 95 & 0\\
            ARGO & all & 0.35 & 95 & 0\\
            \hline
        \end{tabular}
        \caption{Summary of the properties of the detectors considered in this work. The detected neutrino flavor is the flavor associated with the with the dominant interaction for $\mathcal{O}(10~\mathrm{MeV})$ neutrinos at each experiment. The efficiency at $20$~MeV is the average fraction of interactions detected for a $20$~MeV neutrino. Finally, background rates due to PMT noise and radioactivity are taken into account for IceCube and KM3NeT and neglected for all other experiments. The CCSN event selection cuts are harsher at KM3NeT than at IceCube due to the higher rates of radioactive decays in seawater.}
        \label{tab:experiments}
  \end{center}
    \end{table*}
\subsubsection{Water Cherenkov detectors: KM3NeT, IceCube, Hyper-Kamiokande}
Water Cherenkov detectors are primarily sensitive to CCSN electron antineutrinos.
Current and upcoming experiments have large instrumented volumes and hence would detect particularly high numbers of events if a Galactic CCSN occurred --e.g. $\mathcal{O}{(50000)}$ events at the Galactic Center for Hyper-Kamiokande. However, detectors located in natural environments, such as IceCube and KM3NeT, also suffer from sizable, megahertz-scale, backgrounds from ambient radioactivity and PMT noise.
\paragraph{KM3NeT:} Currently under construction and taking data in the Mediterranean Sea, KM3NeT~\cite{km3net} is composed of two large arrays of detection modules. When completed, in 2028, the detector will survey a volume of $1.5~\mathrm{km}^3$ of seawater. While KM3NeT has been originally designed to detect GeV to PeV neutrinos,  it is also expected to be sensitive to more than $95$\% of Galactic CCSNe~\cite{km3net_sn}. KM3NeT therefore runs a CCSN realtime analysis system and is part of SNEWS~\cite{snews}. However, in spite of its large instrumented volume, the sensitivity of the experiment is limited by important backgrounds from bioluminescence and $^{40}\text{K}$ decays in seawater. For this study, we consider KM3NeT's expected final configuration. We use the detection efficiencies estimated for KM3NeT's current CCSN analysis; the background rate after the CCSN event selection if of  $4.5\times 10^6$ events per second. Additionally, event-by-event reconstruction of the CCSN neutrino energies is not possible and only global properties of the time-integrated spectrum can be constrained~\cite{km3net_sn}.
\paragraph{IceCube:} Located at the South Pole, IceCube~\cite{icecube} is a cubic-kilometer neutrino detector consisting of over $5000$ detection modules  located deep within the Antarctic ice sheet. Operational since 2010, it focuses on high-energy neutrino detection, particularly from astrophysical sources like cosmic rays interacting with the atmosphere. IceCube's extensive detection volume of $1~\mathrm{km}^3$ also makes it highly sensitive to low-energy neutrinos from Galactic CCSNe. In particular, the large detection rates expected then would make IceCube uniquely positioned to determine the neutrino mass ordering and characterize the signatures of hydrodynamical instabilities inside the collapsing star~\cite{icecube_sn}. Consequently, IceCube is a major member of SNEWS~\cite{snews}. Like KM3NeT, however, IceCube suffers from large backgrounds due to radioactive activity and PMT noise and event-by-event energy reconstruction is not possible~\cite{icecube_sn}. In what follows, we take IceCube's CCSN neutrino detection efficiencies from SnowGLoBES. The background rate after the CCSN event selection is of $1.5\times10^6$ events per second.~\cite{icecube_sn_proceeding}.   
\paragraph{Hyper-Kamiokande:} The Hyper-Kamiokande neutrino detector~\cite{hk}, currently under construction at Mount Ikeno in Japan, will be a large-scale WC detector  surveying a volume of $260$~kT of ultra-pure water. Aimed at the detection of MeV to GeV-scale neutrinos, this experiment is expected to play a leading role in CCSN neutrino searches. For a CCSN located at the Galactic Center, for example, HK is expected to detect $\mathcal{O}(50000)$ events~\cite{hk_sn}. With a low energy threshold of around $4$~MeV, HK will trigger on most CCSN neutrinos, with an efficiency larger than $99\%$ and extremely low backgrounds. In this analysis, we consider HK's backgrounds for CCSN neutrino detection to be negligible.  
\subsubsection{$\nu_e$ detection experiments: Deep Underground Neutrino Experiment}
The Deep Underground Neutrino Experiment (DUNE)~\cite{dune}, based at the Fermi National Accelerator Laboratory and at the Sanford Underground Research Facility, will be a neutrino beam experiment which primarily aims at measuring the neutrino mass ordering and the CP phase in the lepton sector. However, its far detector, a liquid argon time projection chamber (LArTPC), comprising four $10$~kton modules, can also be used to detect atmospheric and cosmic neutrinos. In the CCSN neutrino energy regime, DUNE would be primarily sensitive to electron neutrinos and is expected to detect $\mathcal{O}(1000)$ events for a CCSN at the Galactic Center, with negligible backgrounds~\cite{dune_sn}. DUNE is expected to be completed at the end of the 2020s and its large scale, low background, and sensitivity to $\nu_e$ will make it play a key role in Galactic supernova searches. In this work, we consider DUNE's backgrounds for CCSN neutrino detection to be negligible. 
\subsubsection{Coherent elastic neutrino-nucleus scattering experiments: DarkSide-20k and ARGO}
DarkSide-20k~\cite{darkside}, located at the Gran Sasso National Laboratory, is designed for the detection of dark matter particle candidates. The detector utilizes a liquid argon time projection chamber (TPC) technology, which detects the ionization electrons and scintillation light generated by the interaction of dark matter particles with argon nuclei. Beyond dark matter searches, DarkSide-20k is sensitive to coherent elastic neutrino-nucleus scattering (CE$\nu$NS) and would detect $\mathcal{O}(100)$ events for a CCSN at the Galactic Center, with negligible backgrounds~\cite{darkside_sn}. Its precursor, DarkSide-50 is part of SNEWS~\cite{snews}. Since CE$\nu$NS treats all flavors equally, DarkSide could provide unique information on the total flux of active neutrinos emitted by a CCSN, although, like IceCube and KM3NeT, it cannot perform event-by-event energy reconstruction. Its main limitation is its small instrumented volume, leading to particularly low detection rates in the first tens of milliseconds of the CCSN. We therefore also consider a similar, next-generation, experimental project, ARGO~\cite{darkside_sn}, which will be 7 times larger than DarkSide-20k. In what follows, we will assume ARGO to differ from DarkSide only by its size. Moreover, we will neglect the backgrounds associated with CCSN neutrino detection for both experiments, as these background were estimated to be more than two orders of magnitude lower than the expected CCSN signal from a $11~M_\odot$ progenitor at $10$~kpc~\cite{darkside_sn}.\\\newline

Spanning a diverse range of capabilities and encompassing all the detectable flavor combinations expected in future-generation experiments, this selection of detectors, while comprehensive, is not exhaustive. In particular, we considered Hyper-Kamiokande rather than Super-Kamiokande as the former is more likely to operate when DUNE and ARGO are also active. Another key experiment for CCSN investigations in the coming decades is the Jiangmen Underground Neutrino Observatory (JUNO). JUNO can indeed measure both electron antineutrinos and the collective sum of all neutrino flavors, effectively serving as two complementary detectors simultaneously~\cite{juno,juno2}. While efficiency curves for JUNO's CCSN detection channels are not currently publicly available, JUNO, once built, is poised to significantly enhance the study of the flavor-dependent behaviors of CCSN neutrinos.

\subsection{Neutrino ﬂavor conversion}
\label{subsec:flavor}
\begin{figure}
   \centering
   \includegraphics[width=0.6\linewidth]{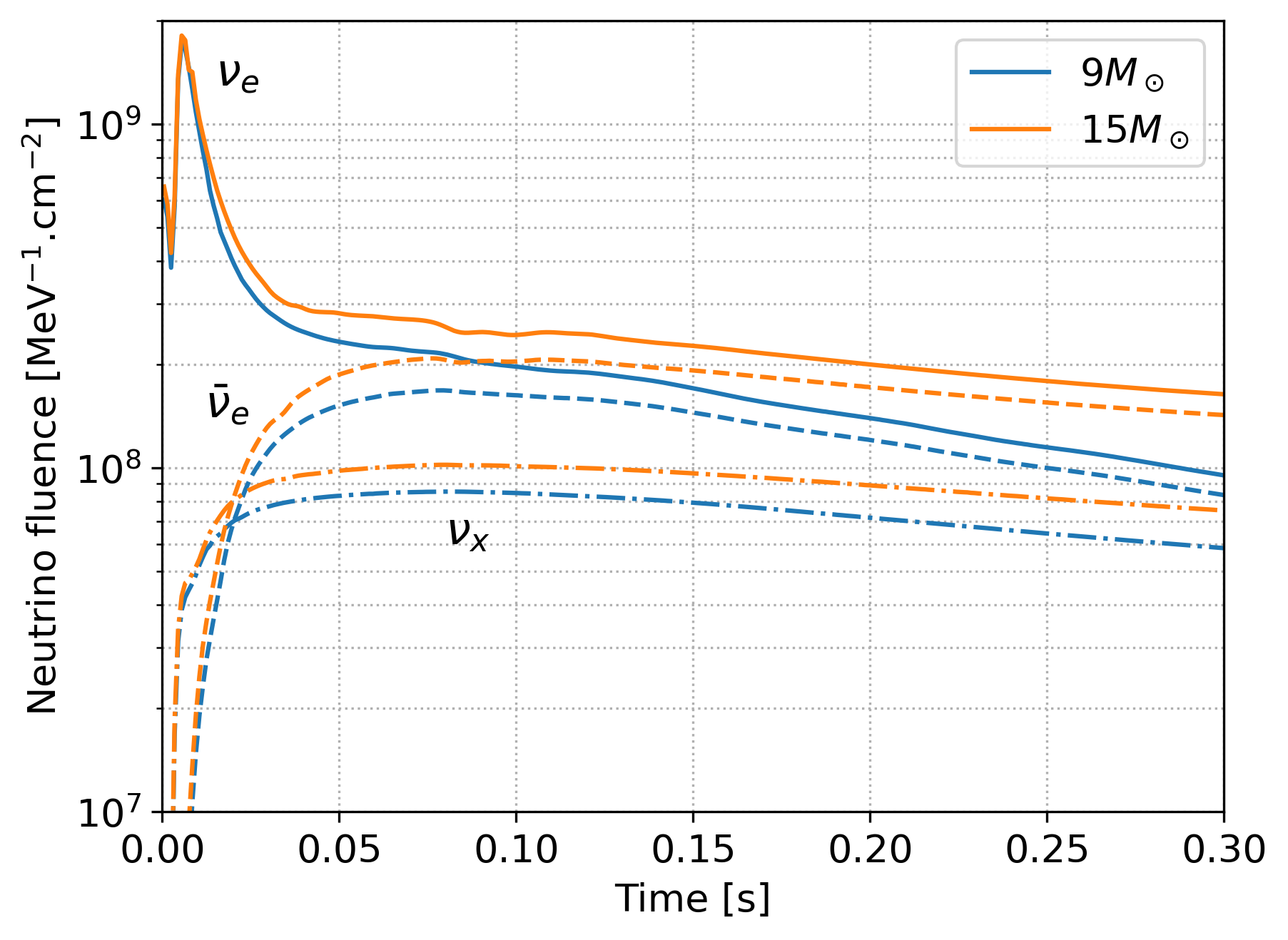}
   \caption{Neutrino luminosities as a function of time for the $9~M_\odot$ and $15~M_\odot$ models from~\cite{segerlund}. The solid lines correspond to electron neutrinos, the dashed lines to electron antineutrinos and the dash-dotted lines to a species representative of heavy-lepton neutrinos and antineutrinos $\nu_x$. The emission peak for $\nu_e$ during the first $30$~ms corresponds to the neutronization phase. After this period, the fluxes for the electron neutrinos and antineutrinos become comparable as the star enters the accretion phase.}
\label{fig:fluences_masses}
 \end{figure}

The initial tens of milliseconds following the core bounce are characterized by an intense emission of electron neutrinos, a period known as ``neutronization''. This phenomenon is associated with the initial propagation of the shock wave through the protoneutron star (PNS) (for a detailed review, see~\cite{janka2012,janka2017,muller2020}). At the same time, the accretion of matter onto the PNS leads to the production of a large number of both electron neutrinos and antineutrinos by charged-current processes in the PNS hot mantle. This phase, known as ``accretion phase'', can last for a few hundreds milliseconds. Neutrino emission profiles for both phases are depicted in Figure~\ref{fig:fluences_masses} for different supernova models. In contrast, muon and tau neutrinos are produced solely by the cooling of the deepest core regions and their emission flux increases at a much slower rate. Therefore, during the first hundreds of milliseconds of a core-collapse supernova, there is a significant disparity in the fluxes of electron neutrinos, electron antineutrinos, and other neutrino types (see Fig.~\ref{fig:fluences_masses}). The observed CCSN neutrino rates on Earth hence strongly depend on the evolution of neutrino flavor composition, both within the star and along their journey to Earth. In this section, we discuss the dominant flavor conversion process in the SM, i.e. the adiabatic Mikheyev-Smirnov-Wolfenstein (MSW)~\cite{adiabaticMSW} transitions, and provide an overview of new physics processes susceptible to modify neutrino rates in the neutronization phase. We focus particularly on neutrino two-body decays, which will be used as a case study for our analysis.

\subsubsection{Adiabatic MSW effect}
\label{sec:MSW}
In the first $\mathcal{O}(100~\mathrm{ms})$ of the CCSN, the dominant neutrino flavor conversion process inside the collapsing star is adiabatic MSW transitions~\cite{adiabaticMSW}. These transitions convert electron neutrinos (antineutrinos) into the heaviest (intermediate) neutrino (antineutrino) mass eigenstate $\nu_h$ ($\bar{\nu}_m$), while muon and tau neutrinos become combinations of the other two eigenstates. The flavor composition of the neutrino fluxes arriving on Earth can hence be expressed as a function of the electron flavor survival probability $p$:
\begin{align}
    \Phi_e^\mathrm{Earth} &= p \Phi_e^{(0)} + (1-p) \Phi_x^{(0)}\\
    \Phi_x^\mathrm{Earth} &= \frac{1-p}{2} \Phi_e^{(0)} + \frac{1+p}{2} \Phi_x^{(0)}
    \label{eq:survival_msw_sm}
\end{align}
where $\Phi^{(0)}$ is the initially produced neutrino flux and $x$ refers to the $\mu$ and $\tau$ flavors. A similar formula holds for antineutrinos using the probability $\bar{p}$. For neutrinos, $p \approx 0.30$ in the inverted mass ordering (IMO) and $0.02$ in the normal mass ordering (NMO). For antineutrinos, $\bar{p} \approx 0.02$ in the IMO and $0.70$ in the NMO~\cite{adiabaticMSW}.

\subsubsection{New physics scenarios: a case study with neutrino two-body decays}
At first glance, the possibilities of BSM scenarios altering the neutronization flux seem limitless. However, BSM physics in the neutrino sector is already tightly constrained by beam, solar, and atmospheric neutrino measurements. In order to distort the neutrino flux sufficiently to mislead CCSN alert systems while still being allowed by current observations, a new physics process needs to be tied to conditions specific to CCSN measurements: the high density of the CCSN core, the possible presence of large magnetic fields with a complex structure, and the particularly long neutrino travel path from the CCSN to the Earth. To date, three types of processes susceptible to substantially modify CCSN neutrino fluxes have been proposed: non-standard neutrino self-interactions~\cite{self-interactions}, the conversion of active neutrinos into sterile neutrinos~\cite{sterile,sterile2,spinflip}, and neutrino two-body decays~\cite{ando_decay,degouvea}. 
While the corresponding new physics models are associated with a wide parameter space, past studies considered simplified frameworks efficiently modelling entire classes of BSM scenarios with only two or three parameters. Since the number of these simplified frameworks is limited by the tight existing constraints on neutrino properties, it is feasible for an alert system to consider several representative new physics models to check for possible departures from the SM. In this work, we assess the impact of such a study using the example of neutrino two-body decays, which allows studying a highly diverse landscape of possible CCSN  neutrino signal distortions.

In the SM, heavy neutrinos species can decay into lighter species only through a loop-level process, with a lifetime larger than the age of the Universe. However, this lifetime can be considerably shortened in simple extensions of the SM where a heavy neutrino mass eigenstate can undergo a two-body decay into a lighter neutrino and a new scalar field. These BSM scenarios have been the focus of thorough investigation and have been constrained by a wide range of measurements, as extensively described in~\cite{degouvea} and \cite{volpe_decay}; the latter, in particular, derived new stringent constraints on certain classes of neutrino decay models. 
Overall, models where at least two neutrino species decay with similar rates have therefore been already strongly constrained up to distance scales of the order of $10$~kpc. However, models featuring significant discrepancies in the lifetimes of neutrino species, particularly with one species having a notably shorter lifetime, remain an intriguing realm of exploration.

In this study, we focus on a scenario discussed in~\cite{degouvea}, where the heaviest neutrino or antineutrino mass eigenstate $\nu_h$ decays into the lightest mass eigenstate $\nu_\ell$ and an invisible scalar $\phi$:
\begin{align}
    \nu_{h,L} \rightarrow \nu_{\ell,L/R} + \phi
    \label{eq:nudecay}
\end{align}
where $L$ and $R$ represent the neutrino helicity. 
We consider the $\phi_0$ model of~\cite{degouvea}, where neutrinos are Dirac particles and lepton number is conserved. In this model, active CCSN (anti)neutrinos decay into either active or sterile (anti)neutrinos, depending on whether the interaction conserves or flips helicity.
We parameterize this model using two quantities. First, we define the \emph{normalized supernova distance} as $\bar{r} = d/r_0$, where $d$ is the supernova distance to Earth and $r_0$ is the characteristic decay length for $10$~MeV neutrinos, defined as
\begin{align}
    r_0 = c\tau \left(\frac{10~\mathrm{MeV}}{m}\right),
\end{align}
with $\tau$ and $m$ being the lifetime and the mass of the heavy neutrino eigenstate, respectively. In the SM, the $\bar{r}$ parameter is equal to $0$. Second, we introduce the \emph{branching ratio of visible decays}, $\zeta$, which represents the fraction of helicity-conserving neutrino decays~\cite{degouvea}. For $\zeta=1$, all heavy CCSN neutrinos decay into light active neutrinos, while for $\zeta=0$ all decay products are undetectable.

\begin{figure*}
    \includegraphics[width=\linewidth]{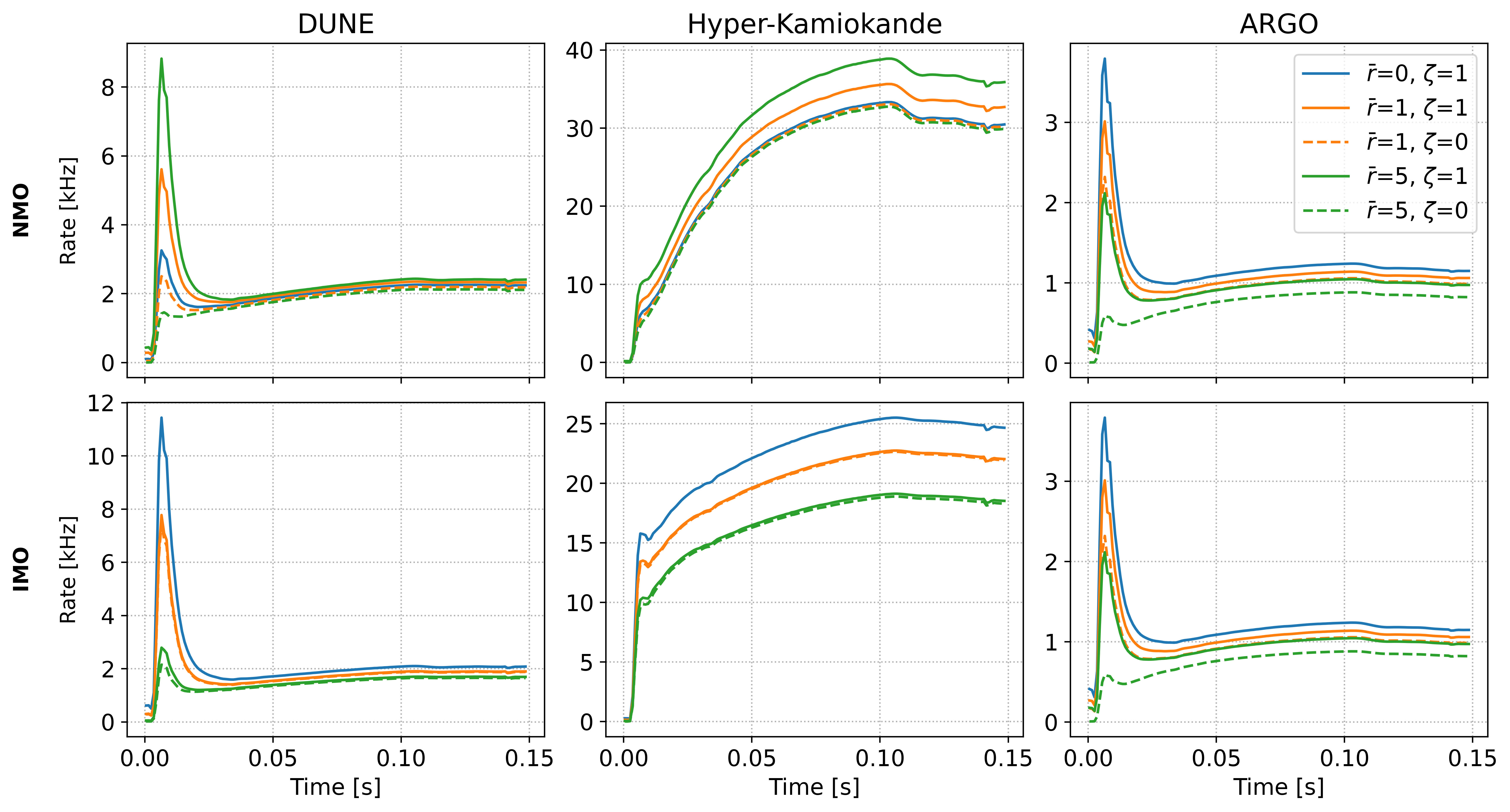}
    \caption{Expected neutrino rates for a $11~\text{M}_\odot$ progenitor~\cite{model} at $10$~kpc at DUNE (left), HK (middle), and ARGO (right), for the normal (top) and inverted (bottom) mass orderings, and for different values of $\bar{r}$ and $\zeta$. The SM case ($\bar{r}=0$) is shown in blue, and decay scenarios with $\bar{r}=1$ and $5$ are shown in orange and green, respectively. Solid (dashed) lines represent models with $\zeta=1$ ($\zeta=0$).}
    \label{fig:decay_snrates}
\end{figure*}

When travelling through the collapsing star, as discussed in Section~\ref{sec:MSW}, adiabatic MSW conversions turn electron neutrinos into the heaviest mass eigenstate $\nu_h$. On their way to Earth, $\nu_h$ ($\bar{\nu}_h$) will partially decay into the lightest active neutrino $\nu_\ell$ ($\bar{\nu}_\ell$), or into sterile neutrinos. The impact of these decays on the neutrino rates during the first tens of milliseconds of the CCSN have been extensively described in~\cite{degouvea}. The variations of the expected neutrino signal time profiles as a function of $\bar{r}$ and $\zeta$ are shown in Figure~\ref{fig:decay_snrates} for different experiments. In what follows, we provide a concise recap of the most significant effects of neutrino decays in the $\phi_0$ model:
\begin{itemize}
    \item \emph{Appearance/disappearance of the $\nu_e$ neutronization peak}: in the NMO, the $\nu_e$ emission peak associated with the neutronization phase is not expected to be visible in $\nu_e$ detection experiments such as DUNE. However, if heavy neutrinos decay into light visible neutrinos, which have a $70$\% probability of being seen as $\nu_e$, this peak can reappear. Conversely, in the IMO, neutrino decays will cause the neutronization peak expected at DUNE to disappear since the electron neutrino component of the lightest species is of only $2\%$. These effects can be seen in the left panels of Figure~\ref{fig:decay_snrates}.
    \item \emph{Increase/decrease of the $\bar{\nu}_e$ rate}: in the NMO, the $\bar{\nu}_e$ rate expected at WC detectors will receive contributions from the visible decays of heavy antineutrinos, driven by the higher likelihood of the lightest neutrino species being identified as $\bar{\nu}_e$. Conversely, the electron antineutrino rate would decrease by up to $30\%$ in the IMO when $\bar{r}$ increases. These effects are illustrated in the middle panel of Figure~\ref{fig:decay_snrates}.
    \item \emph{Decrease of the total neutrino rates}: in the Dirac $\phi_0$ scenario considered here, active neutrinos could decay into sterile neutrinos which would escape detection. Moreover, even for visible decays, the produced neutrinos will be softer and hence less likely to interact. For experiments sensitive to the sum of all neutrino flavors, such as DarkSide-20k or ARGO, the neutrino rate is expected to fall short of the SM predictions. Decays into sterile neutrinos could also lead to a decrease of the $\nu_e$ and $\bar{\nu}_e$ rates expected at DUNE and at WC detectors, regardless of the mass ordering. These effects are shown in Figure~\ref{fig:decay_snrates} for the different detector types. Note that for DUNE and WC detectors, $\zeta$ variations either have a weak impact on detection rates or can lead to degeneracies with the SM. These degeneracies, however, can be lifted by considering experiments sensitive to all neutrino flavors such as DarkSide-20k and ARGO.
\end{itemize}
For sufficiently short neutrino lifetimes, as shown in Figure~\ref{fig:decay_snrates}, the aforementioned effects could potentially alter the expected neutrino rates by several tens of percents, thus substantially impacting CCSN distance measurements. In the NMO and for $\zeta$ close to one, rate increases at DUNE and in WC detectors might lead to an underestimation of the CCSN distance. Nevertheless, this bias could potentially be mitigated by the observation of a neutrino deficit at DarkSide-20k or ARGO. Conversely, within the IMO or for small values of $\zeta$, the decrease of the neutrino rates across all experiments could lead to overestimating the CCSN distance, although the size of this decrease depends on the type of experiment considered. Note that the flux suppression induced by neutrino decays can mimic the one observed in other new physics scenarios such as the ones involving active-sterile neutrino conversions. In either case, a meticulous comparison of neutrino rates among different detector types is imperative to identify deviations from the SM and to make necessary corrections to distance measurements.

\subsection{Analysis pipeline}
\label{subsec:analysis_pipeline}
We generate neutrino ﬂuxes and compute the expected neutrino detection rates using SNEWPY v1.3b1~\cite{snewpy}. Interaction cross-sections, as well as the detection efficiencies and energy reconstruction effects, are modelled using the SnowGLoBEs software~\cite{snowglobes}. For DarkSide-20k and ARGO, which have not been included in SnowGLoBEs yet, we use the detection efficiencies from~\cite{darkside_sn}. For IceCube and KM3NeT, where radioactivity and PMT noise cannot be neglected, we add Poissonian backgrounds to the signal, with constant rates of $1.5$~MHz~\cite{icecube_sn_proceeding} and $3$~MHz~\cite{km3net_sn}, respectively. Finally, we modified the SNEWPY public version to include the neutrino decay model discussed in Section~\ref{subsec:flavor}. Semi-analytical expressions for the neutrino fluxes expected on Earth for these models are given in Appendix~\ref{appendix:decays}.

For each CCSN model and neutrino detector considered, expected neutrino rates are evaluated both in the NMO and the IMO, for the following $(\bar r,\zeta)$ values:
\begin{align}
    \bar{r} &= \begin{cases}
        0.1k \quad\text{for }k\in \llbracket 0~;~ 9 \rrbracket\\
         k \quad\text{for }k\in \llbracket 1~;~ 10 \rrbracket
    \end{cases}
    \label{eq:decay_grid}
    \\
    \nonumber
    \zeta &= 0.1k \quad\text{for }k\in \llbracket 0~;~ 10 \rrbracket
\end{align}
For $\bar r=10$ all heavy neutrinos and antineutrinos will have decayed before reaching Earth, and therefore we do not study models with larger $\bar r$ values. To evaluate the neutrino rates for any value of $\bar{r}$ and $\zeta$, we then approximate the $(\bar{r},\zeta)$ dependence of the CCSN rate by a polynomial function, as described in Appendix~\ref{appendix:interpolation}.

\section{Constrain all CCSN properties at once}
\label{sec:likelihood_approach}
In this section, we present a likelihood-based analysis framework which could be used by a CCSN alert system to simultaneously estimate the CCSN distance and the properties of the neutrinos, by fitting the neutrino event counts in suitably chosen time windows for multiple experiments. 
\label{sec:ccsn_full}

\subsection{Choosing time windows}
\label{subsec:observables}
\begin{figure*}
   \centering
   \includegraphics[width=0.56\linewidth]{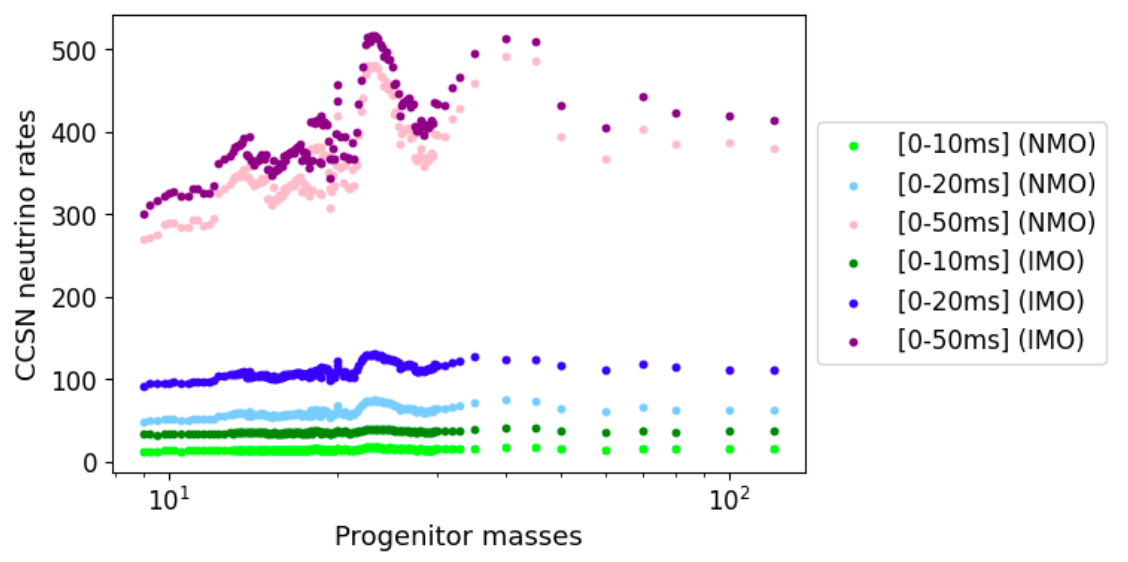}
   \includegraphics[width=0.42\linewidth]{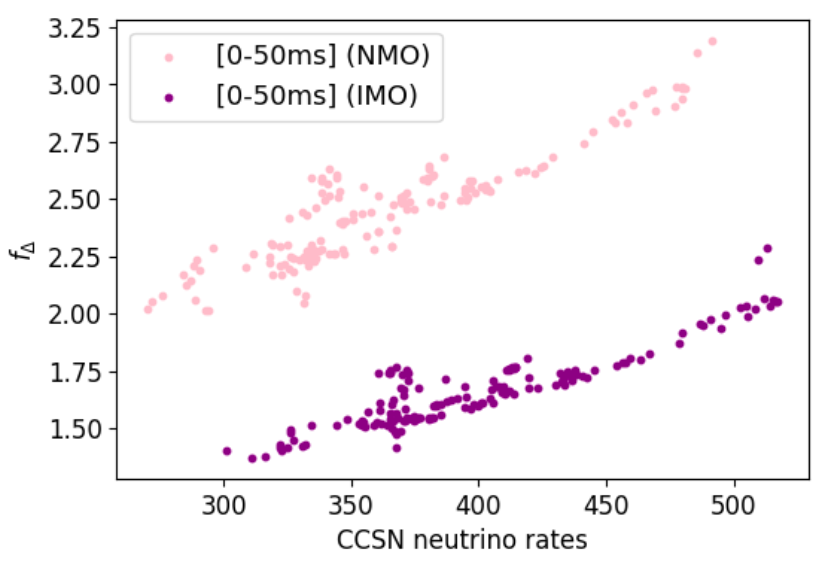}
   \caption{Left: CCSN neutrino event counts in the Hyper-Kamiokande detector for a CCSN at $10$~kpc as a function of progenitor masses, for the first $10$~ms (green), $20$~ms (blue), and $50$~ms (purple) postbounce. Right: $f_\Delta$
as a function of the number of CCSN neutrino events within the first $50$~ms after bounce. For both panels, the light and dark shades represent the NMO and the IMO, respectively. Note that, for the first $20$~ms postbounce, the NMO and IMO predictions do not overlap, even when no assumption is made about the CCSN model. For the first $50$~ms, however, the IMO and NMO scenarios can be distinguished only if the range of possible CCSN models is constrained.}
\label{fig:rates_hk}
 \end{figure*}

We base the choice of time windows for this study on two categories of observables, which have been shown in the literature to either have a weak dependence on the CCSN model, or to be easily parameterizable as a function of the supernova progenitor properties.
\paragraph{Neutrino rates in the neutronization phase:} the neutrino rates measured during the first tens of milliseconds postbounce have been shown to only weakly depend on the stellar progenitor properties~\cite{kachelriess,oconnor}. This model dependence is expected to be minimal when considering the first $10$~ms after the onset of the CCSN neutrino signal, a time region which contains the bulk of the $\nu_e$ neutronization peak. However, for most experiments, the number of events expected in this narrow window is particularly small for most Galactic supernovae and, hence, to reduce statistical uncertainties, longer time periods should be considered. Using the fluxes measured up to $18$~ms postbounce, Kachelriess \emph{et al}~\cite{kachelriess} showed that the CCSN distance could be measured with a precision of $6\%$ in a hypothetical megaton-scale WC detector. For smaller detectors, enlarging this window up to $50$~ms has been proposed by Segerlund \emph{et al}~\cite{segerlund} in the context of analyses at SNEWS. The expected signal in this larger window, however, would significantly depend on the CCSN type, as illustrated in Figure~\ref{fig:rates_hk} (left) for the case of Hyper-Kamiokande.
\paragraph{Neutrino rate ratio f$_\Delta$:} in~\cite{segerlund}, a new observable, $f_\Delta$, was defined to mitigate the residual model dependence of the $N(0-50~\mathrm{ms})$ rates. This observable is defined as the ratio of neutrino rates between the accretion and the neutronization phases:
    \begin{equation}
        f_\Delta= \frac{N(100-150~\mathrm{ms})}{N(0-50~\mathrm{ms})}
        \label{eq:fdelta_orig}
    \end{equation}
    and its dependence on $N(0-50~\mathrm{ms})$ is close to linear, as shown in Figure~\ref{fig:rates_hk} (right). Exploiting this quasi-linear dependence, Segerlund \emph{et al}~\cite{segerlund} proposed an algorithm to infer the CCSN distance from a $(f_\Delta, N(0-50~\mathrm{ms}))$ measurement at a single detector. Accounting for both variations between CCSN models and the CCSN distance distribution in the galaxy, the associated distance measurement uncertainties are of around $5\%$ at HK and IceCube and around $8\%$ at DUNE.   

Based on the observables and studies described above, we selected four time windows as follows. First, exploiting the results from~\cite{kachelriess}, we split the $50$~ms postbounce time interval into three windows: $[0,10]$, $[10,20]$, and $[20,50]$~ms. As a fourth window, following the methodology detailed in~\cite{segerlund}, we choose the $[100,150]$~ms time interval. This choice of time windows allows considering the whole $50$~ms neutronization period while resolving the neutronization peak, for experiments sensitive to $\nu_e$, and mitigating the CCSN model dependence.

 
\subsection{Likelihood analysis}
\label{subsec:likelihood} 
For a given set of neutrino detectors taking data at the time of the CCSN, we define the following likelihood function:
    \begin{align}
        \log\mathcal{L}(\{\mathcal{O}_\mathrm{obs}\} | d, M, \bar r, \zeta, \mathrm{MO}) = \sum_{i} \log P[N_i(10\text{ms})] &+ \log P[N_i(10-20\text{ms})] \\
        + \log P[N_i(20-50\text{ms})] &+ \log P[N_i(100-150\text{ms})]
        \label{eq:likelihood_sn}
    \end{align}
where the index $i$ refers to each neutrino detector, $d$ and $M$ are the supernova distance and progenitor mass, respectively, and $(\bar r,\zeta)$ are the neutrino decay parameters. $P$ is the Poisson probability distribution to observe a given number of events:
\begin{align}
    P(N_i) = \frac{\lambda_i^{N_i}e^{-\lambda_i}}{N_i!},
    \label{eq:poisson_like}
\end{align}
with $\lambda_i(d, M, \bar r, \zeta, \mathrm{MO})$ being the expected value of $N_i$ for the CCSN and neutrino properties considered. $\{\mathcal{O}_\mathrm{obs}\}$ represents the set of measurements performed at the different detectors. When constraining a given parameter $\theta$, e.g. the CCSN distance, we treat the others as nuisance parameters $\Xi$, and define a profile likelihood:
\begin{align}
    \mathcal{L}_\mathrm{prof}(\{\mathcal{O}_i\}|\theta) &= \mathrm{max}_{\Xi} \mathcal{L}(\{\mathcal{O}_i\} | \theta, \Xi)
\end{align}
which can be used for either parameter fitting or hypothesis testing. Depending on the parameter of interest, the nuisance parameters can be any subset of $(d,M,\bar{r},\zeta)$. Since we are considering a discrete, finite, set of CCSN models, the likelihood optimization over $M$ can be performed by testing each model individually. To optimize the likelihood over $(\bar{r},\zeta)$ we use a regular grid with $\Delta\bar{r}=0.05$ and $\Delta{\zeta}=0.1$. We will confirm in Section~\ref{sec:decay_full} that this grid spacing is far below the resolution of the best-performing detectors considered in this study, and therefore this grid search is exhaustive. Finally, to optimize over the CCSN distance, we exploit the dependence of the number of signal events in the inverse CCSN distance squared, computing the optimal distance either analytically or numerically as described in Appendix~\ref{appendix:distance}.

In the rest of this paper, we will describe how to exploit this profile likelihood method to constrain the CCSN distance, the CCSN models, and the neutrino properties. To investigate the performances of the different detectors, we will evaluate the likelihood $\mathcal{L}$ either for a large number of pseudo-experiments or for a ``typical'' measurement for which the Poisson probabilities $P$ are maximal. In the latter case, we take each observed number of events $N_i$ to the equal to the integer part of the expectation value $\lambda_i$ of its underlying Poisson distribution. For non-integer expectation values, $\lfloor\lambda_i\rfloor$ indeed corresponds to the mode of the associated Poisson distribution.

\section{Breaking degeneracies: neutrino mass ordering}
\label{sec:nmo_full}
In this section, we evaluate the rejection p-value of the IMO (NMO) case assuming the real mass ordering is the NMO (IMO). For this study, we will first consider the SM, then investigate the impact of new physics processes on the detectors' performances using an example BSM scenario. This time, in order to account for uncertainties on the CCSN distance and progenitor properties, we perform hypothesis testing using the likelihood function from equation~\ref{eq:likelihood_sn}. For a given set of observations $\{\mathcal{O}_\mathrm{obs}\}$, we optimize the likelihood over the nuisance parameters for each ordering hypothesis and use the ratio of the resulting likelihoods as a test statistic. This test statistic will be defined as:
\begin{align}
    t(\{\mathcal{O}_\mathrm{obs}\}) = \frac{\mathrm{max}_{d,M,\bar{r},\zeta}\left[\mathcal{L}(\{\mathcal{O}_\mathrm{obs}\}| d, M,\bar{r},\zeta, \mathrm{NMO})\right]}{\mathrm{max}_{d,M,\bar{r},\zeta}\left[\mathcal{L}(\{\mathcal{O}_\mathrm{obs}\}|d, M,\bar{r},\zeta, \mathrm{IMO})\right]}
    \label{eq:test_statistics}
\end{align}
where the likelihoods are optimized over the CCSN distance and the CCSN models, setting $\bar{r}=0$ and $\zeta=1$ if only SM neutrino interactions are assumed. If the possibility of neutrino decays is accounted for, the likelihoods will be also optimized over the $\bar{r}$ and $\zeta$ parameters.

To evaluate the significance associated with a given observation, we use a hybrid Bayesian-Frequentist method which has been introduced by Highland and Cousins~\cite{cousins} and used in the context of LHC studies~\cite{cranmer,cranmer2}. Specifically, to evaluate the probability distribution of $t$  for a given mass ordering hypothesis, we sample possible observations from a prior probability distribution defined as:
\begin{align}
 \pi(\mathcal{O}_\mathrm{obs},\mathrm{MO})  = \int \mathcal{L}(\{\mathcal{O}_\mathrm{obs}\}|d',M',\bar{r}',\zeta',\mathrm{MO})\,\pi_d(d')\,\pi_M(M')\,\pi_{\bar{r}}(\bar{r}')\,\pi_\zeta(\zeta')\;\mathrm{d}d'\,\mathrm{d}M'\,\mathrm{d}\bar{r}'\,\mathrm{d}\zeta'
\end{align}
where the $\pi$ functions represent the prior probabilities of the different parameters. The use of priors allows estimating the distribution of the test statistics $t$ before an observation is made. Since this step can be particularly computationally intensive, the use of the hybrid Bayesian-Frequentist method could allow an alert system to get a quick first estimate of the neutrino mass ordering less than a few seconds after a CCSN observation. 

For the CCSN distance, we take a uniform prior in $[0.5,60]$~kpc --thus accounting for CCSNe both in the Milky Way and in the Large Magellanic Cloud. Since the CCSN distance distribution is expected to peak near the Galactic Center~\cite{beacom} and no supernova is expected between the Milky Way and the LMC, the results presented here are expected to be conservative, with the experiments' reach being underestimated. For the CCSN model, we take a prior proportional to the Salpeter initial mass function $\pi(M)\propto M^{-2.35}$, as discussed in Section~\ref{subsec:models}. If only SM processes are assumed, we set $\bar{r}$ and $\zeta$ to $0$ and $1$, respectively. Conversely, when accounting for neutrino decays we choose flat priors on $\bar{r}$ and $\zeta$ with $\bar{r} \in [0,10]$ and $\zeta\in [0,1]$ ---these priors could later be updated by alert systems in collaboration with theorists. 

To assess the performance of various detectors or combinations of experiments, we determine the CCSN distance at which the median p-value  under the IMO hypothesis exceeds $3\sigma$ when the null hypothesis is the NMO, and vice versa. Here, in order to obtain conservative estimates of this $3\sigma$ distance horizon, we consider a $9~M_\odot$ progenitor. Indeed, this progenitor type is associated with both the highest prior probability and the lowest neutrino fluxes in our study.

\subsection{Standard Model case}
\begin{figure*}
    \centering
    \includegraphics[width=0.45\linewidth]{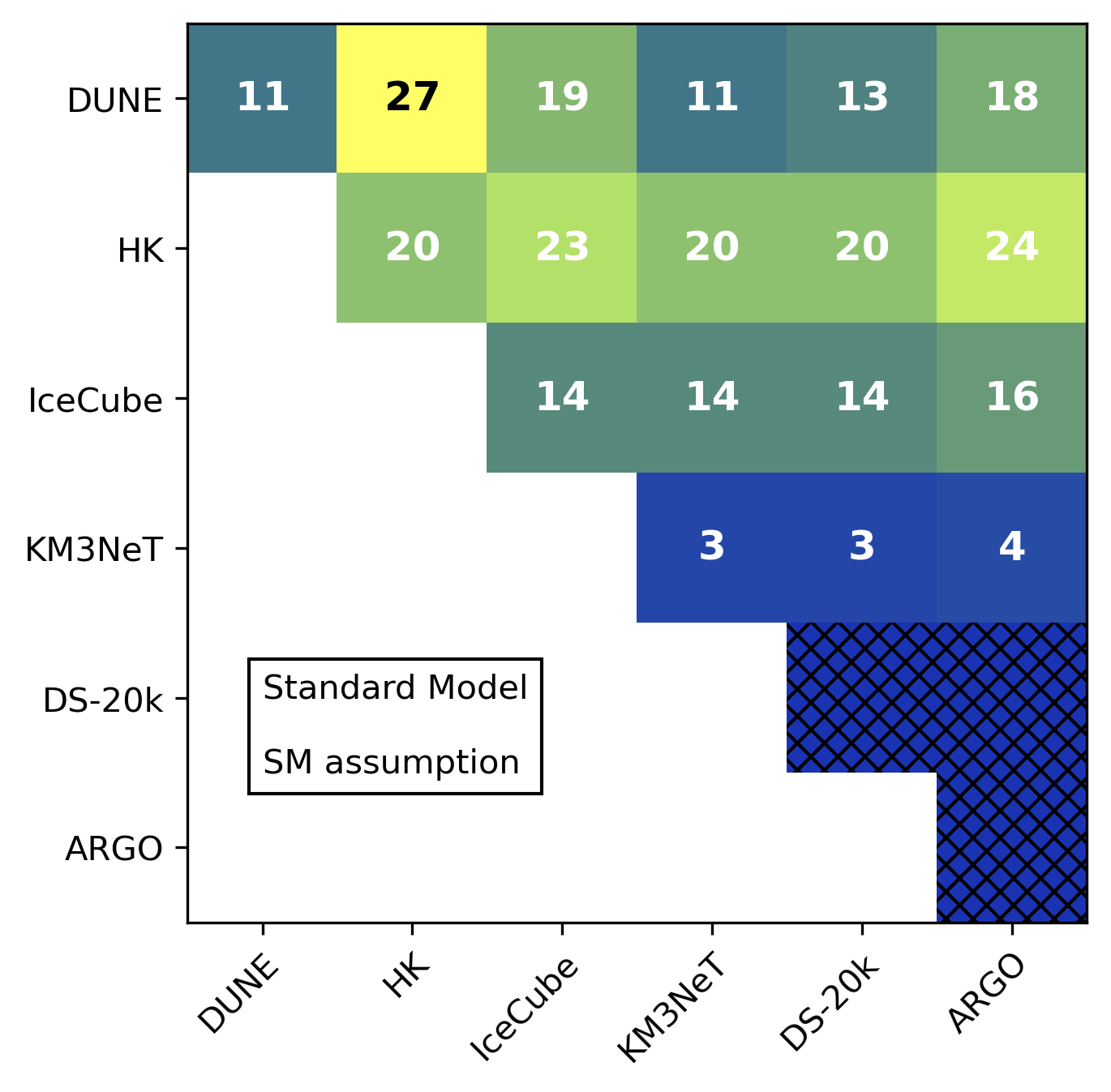}
    \includegraphics[width=0.45\linewidth]{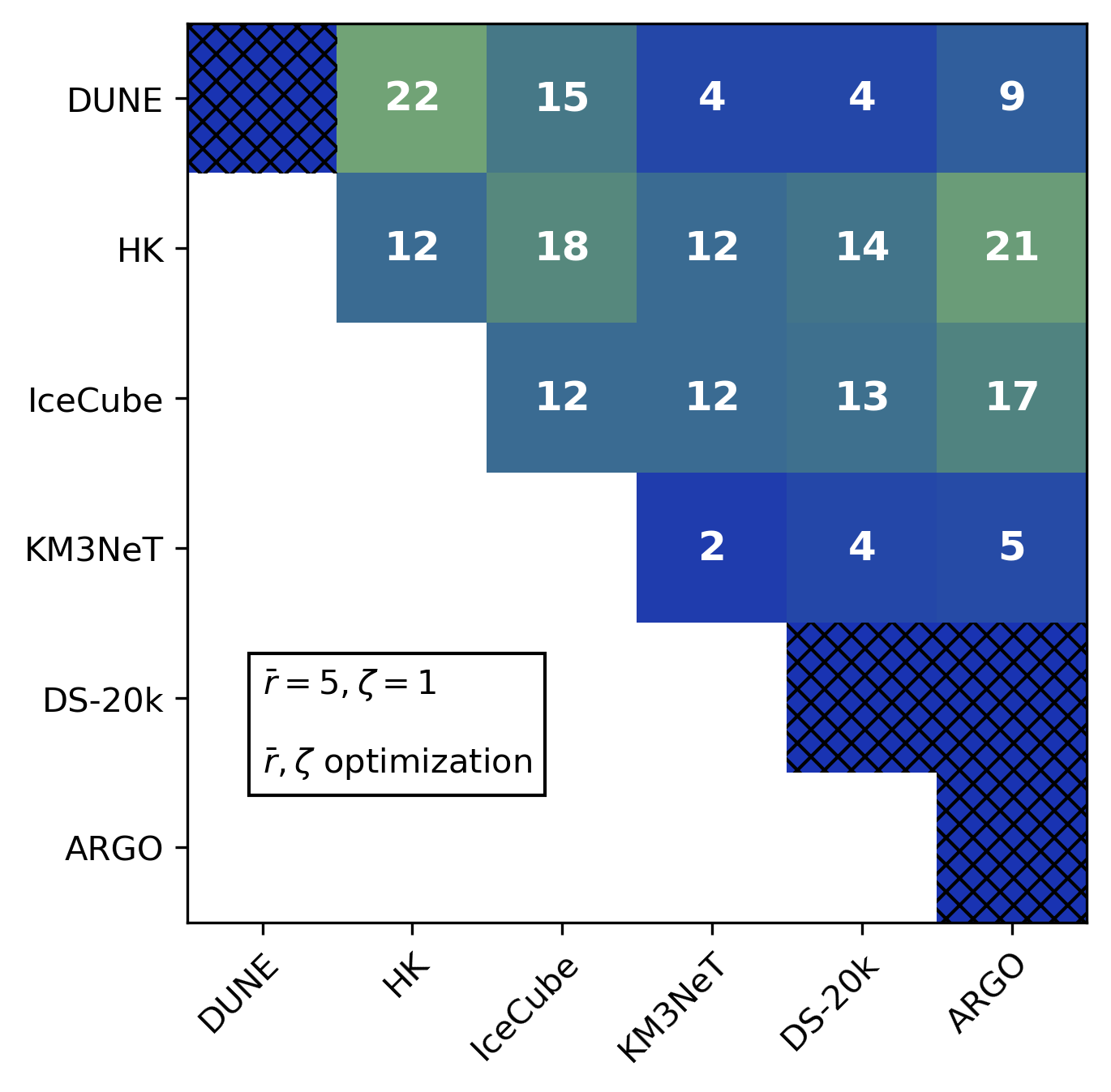}
    \caption{$3\sigma$ distance horizon (in kpc) for rejecting the IMO when the NMO is the true mass ordering. The diagonal represents individual detectors while off-diagonal elements show detector pairs. The hatched regions show experiments or pairs of experiments with no sensitivity to the mass ordering, such as DUNE in the presence of neutrino decays and experiments targeting CE$\nu$NS such as DarkSide-20k and ARGO. Left: SM scenario with a SM likelihood. Right: BSM scenario with $\bar r=5, \zeta=1$, optimizing over $\bar r$ and $\zeta$ in the likelihood.}
    \label{fig:IMOreject_pairs}
\end{figure*}

Figure~\ref{fig:IMOreject_pairs} shows the $3\sigma$ distance horizon for rejecting the IMO hypothesis for individual experiments and for pairs of detectors. We see that pairing Hyper-Kamiokande and IceCube, two large-scale WC detectors, increases this horizon from $20$~kpc (at HK alone) to $23$~kpc. However, the largest increase is obtained when pairing HK with DUNE, with a $27$~kpc horizon covering the whole galaxy. Similarly, the IMO $3\sigma$ distance horizon increases from $11$ to $18$~kpc and $14$ to $19$~kpc when pairing DUNE with ARGO and IceCube, respectively. 
Finally, note that combining KM3NeT or DarkSide-20k with another detector leads to no increase of the horizon. Indeed, due to the small size of DarkSide-20k, only a few events are expected in each bin for CCSNe beyond $10$~kpc at this experiment~\cite{darkside_sn}, while the high KM3NeT noise rates hinder signal characterization beyond a few kpc~\cite{km3net_sn}. 

\begin{figure*}
    \centering
    \includegraphics[width=0.45\linewidth]{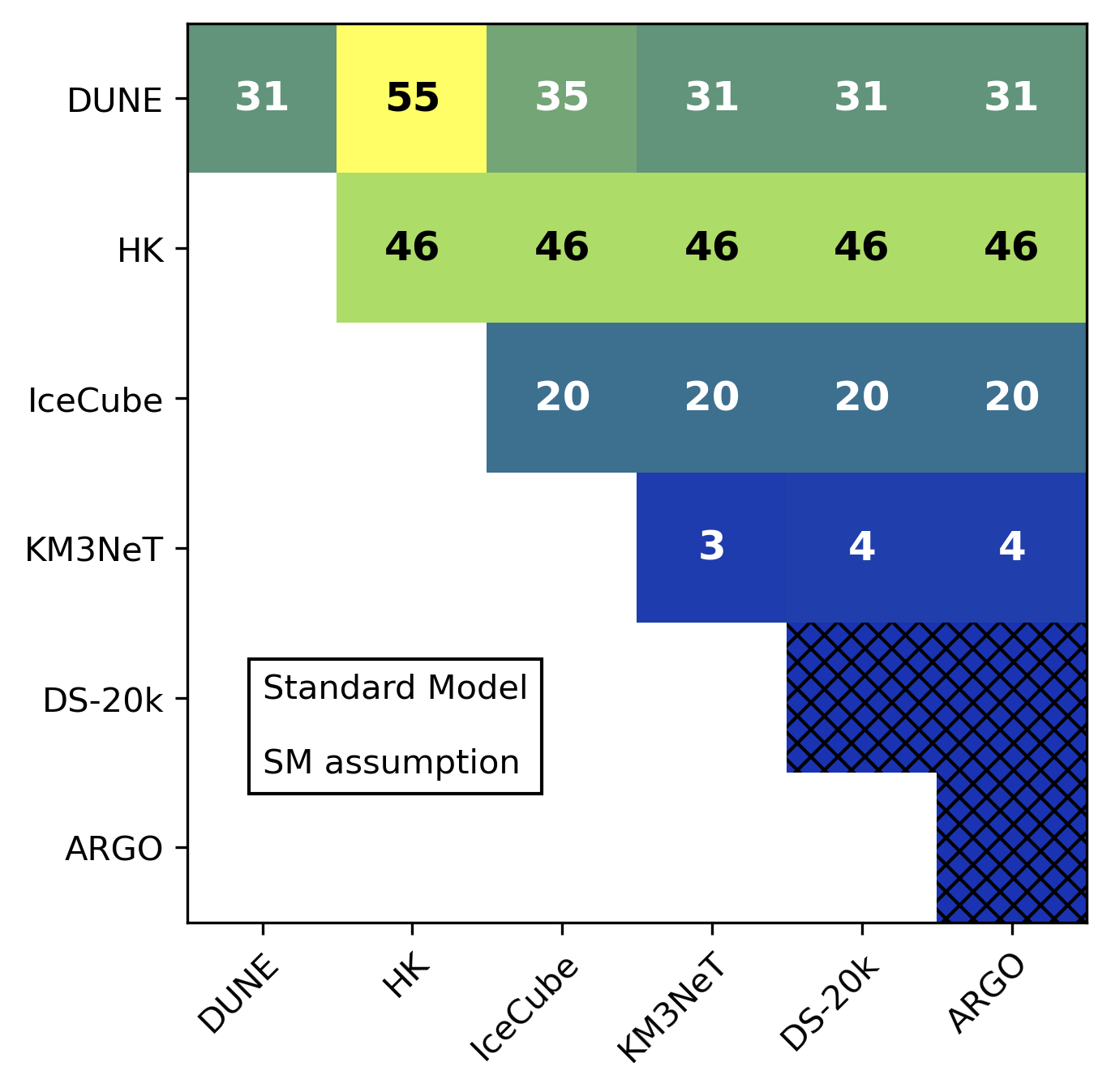}
    \includegraphics[width=0.45\linewidth]{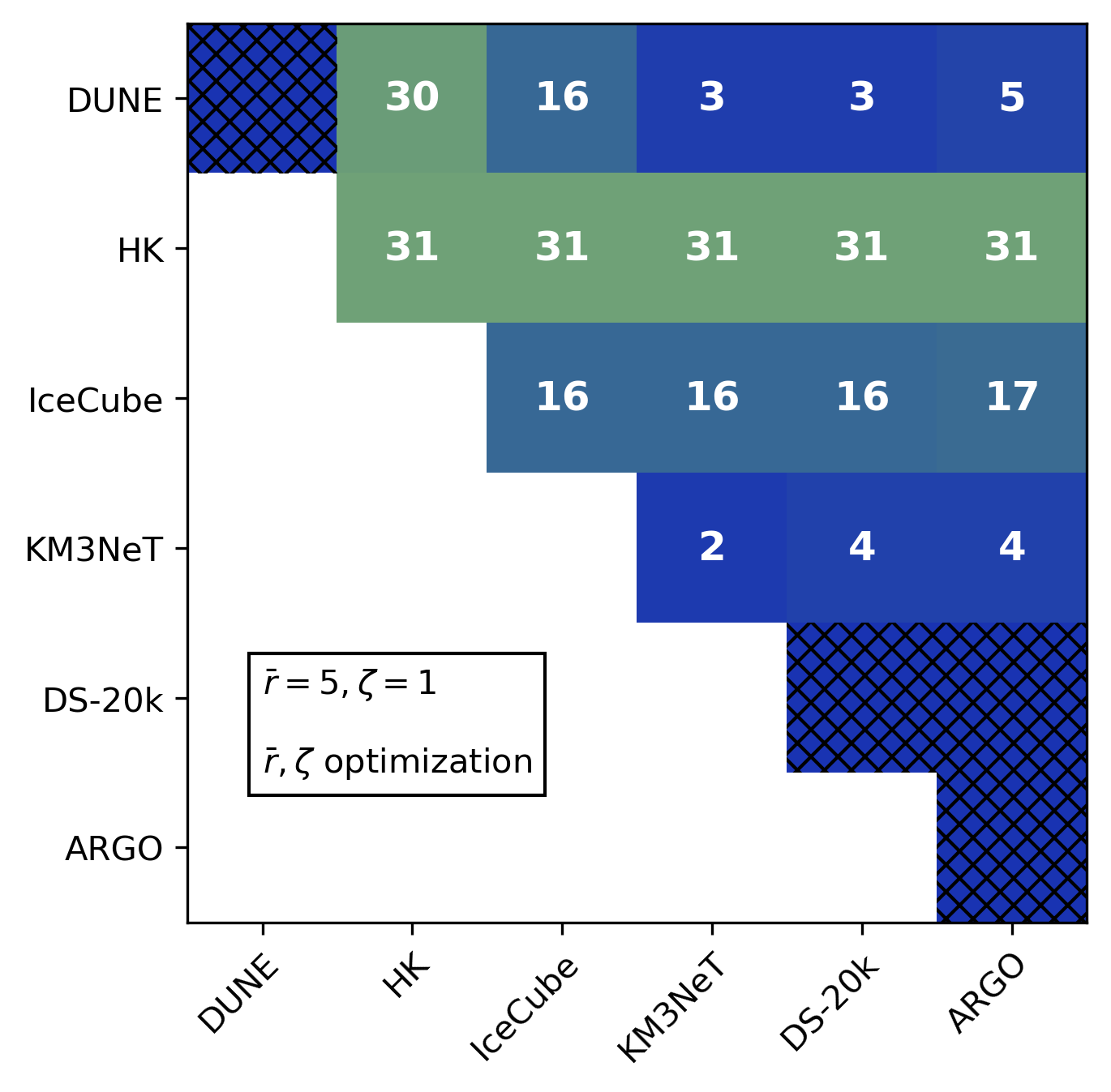}
    \caption{$3\sigma$ distance horizon (in kpc) for rejecting the NMO when the IMO is the true mass ordering. The diagonal represents individual detectors while off-diagonal elements show detector pairs. The hatched regions show experiments or pairs of experiments with no sensitivity to the mass ordering, such as DUNE in the presence of neutrino decays and experiments targeting CE$\nu$NS such as DarkSide-20k and ARGO. Left: SM scenario with a SM likelihood. Right: BSM scenario with $\bar r=5, \zeta=1$, optimizing over $\bar r$ and $\zeta$ in the likelihood.}
    \label{fig:NMOreject_pairs}
\end{figure*}

The $3\sigma$ distance horizons for NMO rejection are shown in Figure~\ref{fig:NMOreject_pairs}. Here, similarly to the IMO rejection case, combining DUNE and HK leads to the highest significances, with a distance horizon extending up to the LMC. 
Combining DUNE with IceCube also leads to a mild improvement of the distance horizon, from $31$ to $35$~kpc. Note that, while these results demonstrate the advantage of combining flavor-complementary neutrino experiments for mass ordering studies, the distance horizons obtained for the NMO rejection scenario will likely shrink when considering more realistic detector models, with non-negligible background rates. Such a study, however, is beyond the scope of this paper since searches for distant supernovae require dedicated analyses (see for example~\cite{sk_mori} at Super-Kamiokande) which have not been designed yet for the experiments we consider.

\subsection{Impact of neutrino decays on mass ordering determination}
To assess the impact of the presence of neutrino decays on the determination of the mass ordering, we consider a CCSN with a $9~M_\odot$ progenitor and a neutrino decay scenario with $\bar{r}=5$ and $\zeta=1$. Here again, to evaluate the rejection significance of the NMO (IMO) hypothesis, we consider the median p-value of possible IMO (NMO) measurements. We first consider a situation where, in spite of the presence of neutrino decays, we optimize the likelihoods from equation~\ref{eq:test_statistics} assuming the SM. Due to this (wrong) SM assumption, the distance horizons at the best-performing experiments shrink significantly. For individual WC detectors, the distance horizons for IMO rejection shrink down to $\sim 10$~kpc for HK and IceCube and $2$~kpc for KM3NeT. Combining HK and IceCube leads to a horizon of $17$~kpc, while, in the SM, the reach of this detector pair extended to $23$~kpc. A similar behavior is observed for NMO rejection. WC detectors retain partial discriminating power, with horizons of $36$~kpc and $16$~kpc at HK and IceCube, respectively. HK therefore retains its ability to probe the mass ordering for any CCSN in the Milky Way. On the other hand, DUNE's capability to reject either mass ordering is completely lost regardless of the mass ordering hypothesis. DUNE's poor performance in this context arises from significant distortions in the expected electron neutrino rates due to neutrino decays. These distortions can lead to observations in the NMO that resemble a Standard Model IMO scenario, and vice versa. 

We now evaluate the capability of neutrino experiments to distinguish the NMO and IMO scenarios in the presence of neutrino decays when the likelihoods in Equation~\ref{eq:test_statistics} are optimized over $\bar r$ and $\zeta$ in addition to the other parameters. The right panels of Figures~\ref{fig:IMOreject_pairs} and \ref{fig:NMOreject_pairs} show the associated distance horizons for rejecting the SM IMO and NMO hypotheses, respectively. Here again, we consider a measurement corresponding to a neutrino decay scenario with $\bar r=5,\zeta=1$ but we now allow $\bar r$ and $\zeta$ to vary when optimizing likelihoods. For NMO rejection, no sizable increase of the distance horizons are obtained for pairs of experiments compared to single detectors. The slight shrinking of the horizon observed for HK+DUNE compared to HK alone is due to our choice of priors\footnote{In DUNE, when only event counts are considered, the IMO and NMO cannot be distinguished if $\bar{r}$ and $\zeta$ are unknown. Due to this degeneracy and to our choice of priors, DUNE's test-statistics distribution for NMO measurements is slightly biased towards ``IMO-like'' (larger than $1$) values. HK observations are not sufficient to correct this bias.}, which does not affect the interpretation of the results since no CCSN is expected in the $25-50$~kpc region. Conversely, for  IMO rejection the highest distance horizons are obtained by pairing DUNE with HK ($22$~kpc) and HK with ARGO ($21$~kpc). These distance horizons include more than $99\%$ of the CCSN candidates in the Milky Way~\cite{beacom}.

The results obtained in this section demonstrate the capability of most future experiments to  discriminate neutrino mass ordering hypotheses for more than $99\%$ of the CCSN candidates in the Milky Way, even when new degrees of freedom allowing departures from SM physics are introduced. Moreover, several of the experiments considered in this analysis, such as DUNE, KM3NeT, and HK, also aim at measuring the neutrino mass ordering using neutrino beams or atmospheric neutrinos~\cite{dune,km3net,hk}. In the rest of this paper, when estimating other parameters, we assume the mass ordering to be known unless stated otherwise. Additionally, due to the limited performance of KM3NeT and DarkSide-20k, in the following sections we will focus on the DUNE, HK, IceCube, and ARGO experiments.

\section{Supernova distance measurement}
\label{sec:distance}
If a nearby CCSN occurs, the neutrino burst will be detected minutes to hours before the electromagnetic emission~\cite{ccsn_time}. It is therefore essential for neutrino detectors to accurately locate the CCSN in real time and transmit this information to telescopes. Estimating the CCSN distance is all the more necessary when the supernova takes place in the Galactic plane, where dust can hinder optical measurements~\cite{distance_dust}. However, distance measurement algorithms proposed in the literature assume that neutrino properties are given by the SM~\cite{kachelriess,segerlund}. In this section, we evaluate the bias in CCSN distance estimates introduced by ignoring neutrino decays for the $\phi_0$ scenario introduced in de Gouvêa \emph{et al}~\cite{degouvea} and discussed in Section~\ref{subsec:flavor}. 

Up to now, the methods proposed in the literature~\cite{kachelriess,segerlund}, and considered by SNEWS~\cite{snews}, to evaluate the CCSN distance make use of the measured CCSN neutrino rates during the first $50$~ms following the supernova observation at a given detector. One approach also incorporates information from the early accretion phase --the first $100$ to $150$~ms after the CCSN trigger-- to reduce the dependence in the progenitor properties~\cite{segerlund}. Here, to estimate the CCSN distance and the associated uncertainties, we maximize the likelihood shown in equation~\ref{eq:likelihood_sn} over the CCSN and neutrino model parameters --assuming the mass ordering is known-- and study its variations as a function of the CCSN distance. 

\subsection{Methodology: distance fit and uncertainty evaluation}
We consider two possible assumptions for the likelihood optimization: the SM assumption, where $\bar r$ is set to $0$ and $\zeta$ to $1$, and the BSM assumption where $\bar r$ and $\zeta$ are allowed to vary. These scenarios are described by the following likelihood profiles:
\begin{align}
    &\mathcal{L}_\mathrm{BSM}(d,\mathrm{MO}) = \mathrm{max}_{\bar{r},\zeta,M}\mathcal{L}(d,M,\bar r,\zeta,\mathrm{MO}) \\
    &\mathcal{L}_\mathrm{SM}(d,\mathrm{MO}) = \mathrm{max}_{M}\mathcal{L}(d,M,\bar r=0,\zeta=1,\mathrm{MO})
    \label{eq:Ldistance}
\end{align}
For either assumption, we consider a typical measurement, as defined in Section~\ref{subsec:likelihood} for a given set of CCSN and neutrino parameters and evaluate the corresponding likelihood profile for different CCSN distances. We then evaluate the confidence interval on the CCSN distance using Wilks' theorem.

\subsection{Results}
\begin{figure*}
    \centering    \includegraphics[width=0.6\linewidth]{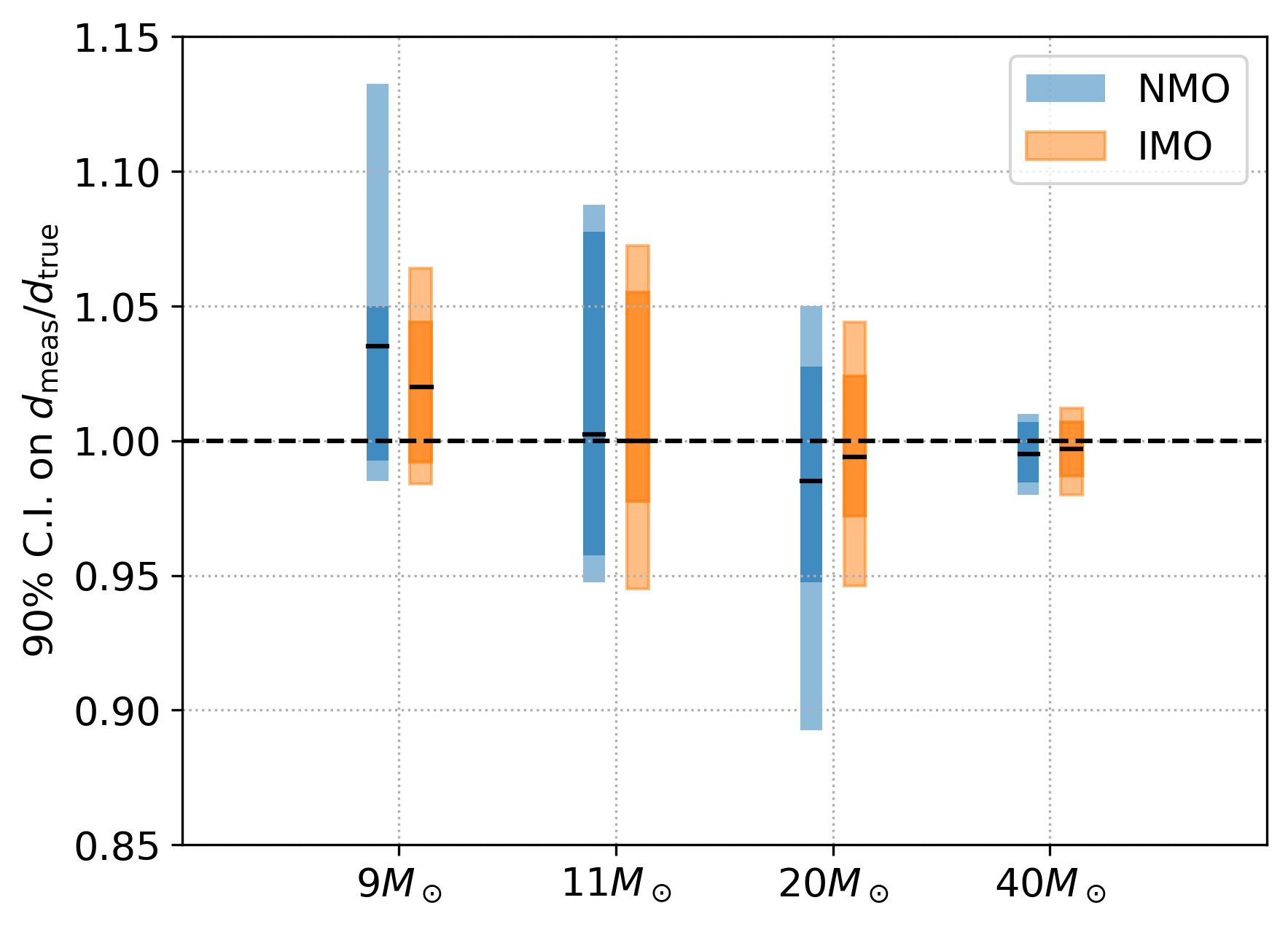}
    \caption{$90\%$ and $68\%$ confidence interval on the measured CCSN distance, normalized by the true distance, for a supernova at 10~kpc. The measurement is made by combining DUNE and Hyper-Kamiokande. Here, CCSN models with progenitor masses of $9$, $11$, $20$, and $40$ solar masses are considered. The black lines show the median distances. The confidence interval always contains the true CCSN distance. Neutrino properties are described by the SM.}
    \label{fig:CLdistance_mass}
\end{figure*}

 We first evaluate the impact of the CCSN model choice on distance measurements by estimating the CCSN distance for four progenitors with $9M_\odot$, $11M_\odot$, $20M_\odot$, and $40M_\odot$. The first two models feature typical, light, progenitors while the last two involve heavier progenitors with neutrino fluxes up to $\sim100\%$ larger.
 Figure~\ref{fig:CLdistance_mass} shows the $90\%$ and $68\%$ confidence intervals normalized by the true distance for these progenitors at $10$~kpc, for measurements made at DUNE and HK. The size and location of the confidence interval vary significantly between models. In particular, the confidence intervals are skewed towards higher distances for the lighter models, associated with lower rates, and towards lower distances for the heavier models. However, the $68\%$ confidence intervals always contain the true distance value, thus demonstrating the robustness of our fit. This robustness has also been verified for other detector pairs and for individual experiments. In what follows, we will hence focus on a single model,  the $11~M_\odot$ progenitor, which is associated with both a high probability and a small bias. 
 
\begin{figure*}
    \includegraphics[width=\linewidth]{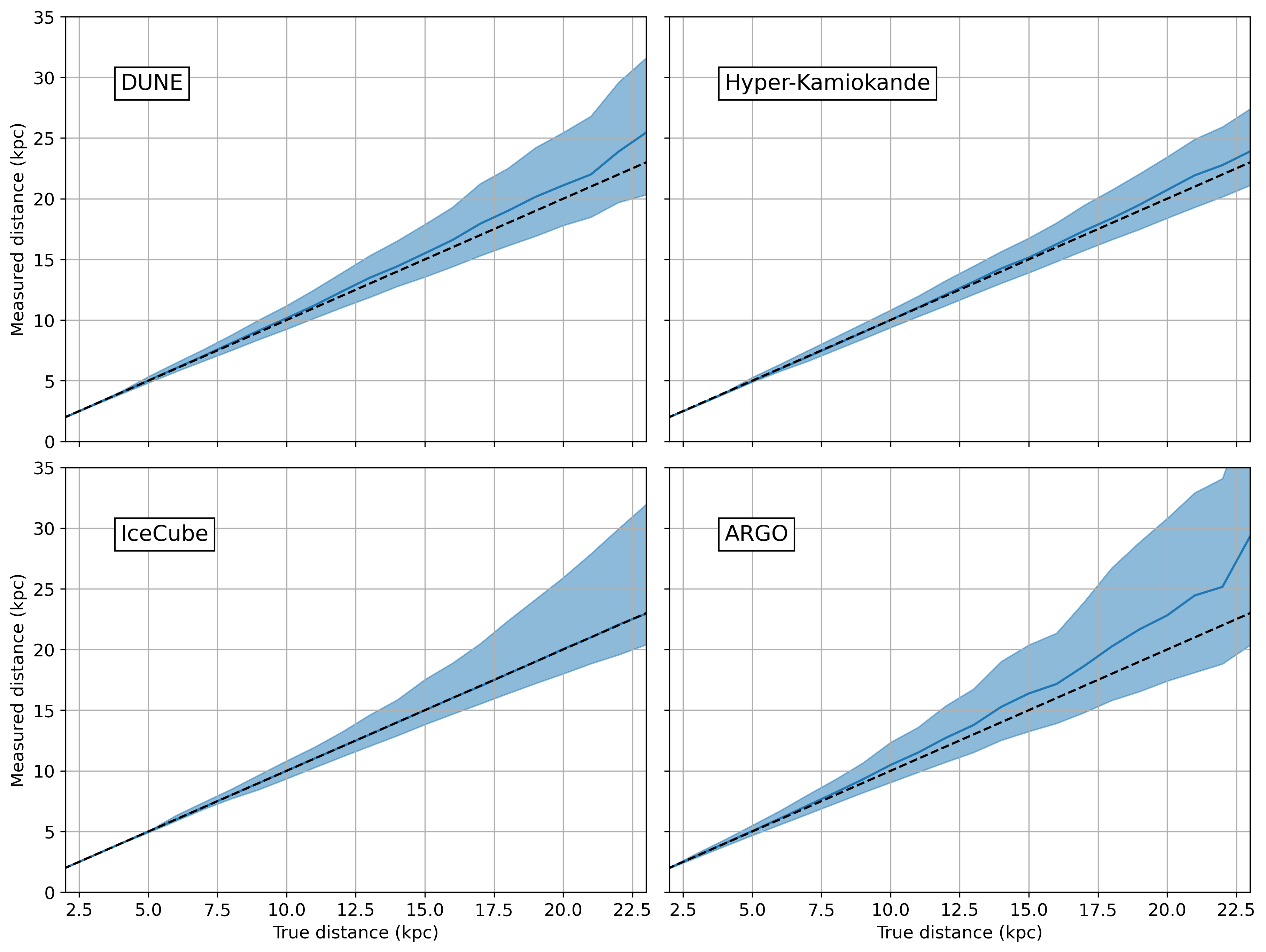}
    \caption{Median value and $90\%$ confidence interval for the measured CCSN distance at DUNE (top left), IceCube (top right), Hyper-Kamiokande (bottom left), and ARGO (bottom right), using the SM likelihood from equation~\ref{eq:Ldistance}, assuming the mass ordering is known to be the NMO and considering a $11~M_\odot$ progenitor. The dashed line represents the true CCSN distance.}
    \label{fig:distance_1detector}
\end{figure*}

\begin{figure*}
    \centering    \includegraphics[width=\linewidth]{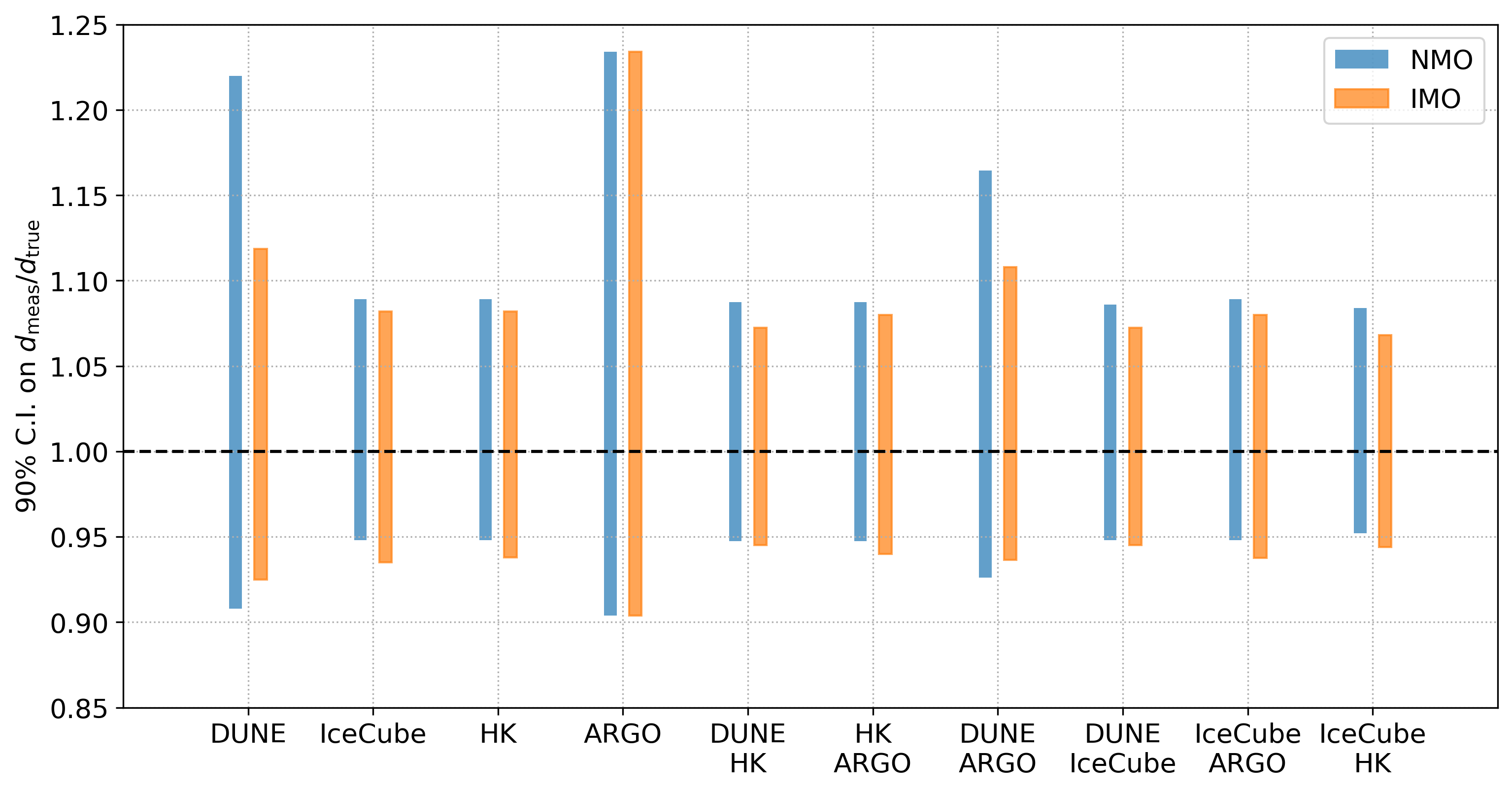}
    \caption{$90\%$ confidence interval on the measured CCSN distance, normalized by the true distance, for a supernova with a $11M_\odot$ progenitor at $10$~kpc. Here, the performances of individual experiments and of pairs of detectors are compared. Neutrino properties are described by the SM.}
    \label{fig:CLdistance_SM}
\end{figure*}

We first consider SM measurements under the SM-only assumption. Figure~\ref{fig:distance_1detector} shows the median value and the $90\%$ confidence interval for the measured distance as a function of the true CCSN distance for the four best-performing experiments from Section~\ref{sec:nmo_full}, under the NMO. At $10$~kpc, the width of the $90\%$ confidence interval ranges from about $1.5$~kpc at HK and IceCube to $3.5$~kpc at ARGO. At $25$~kpc, the furthest edge of the galaxy, this width ranges from $6$~kpc at HK to more than $15$~kpc at ARGO. For the latter, however, distance estimates are biased towards larger values; hence, setting an upper bound of $25$~kpc on the CCSN distance would significantly shrink the confidence range. For a CCSN at $10$~kpc, we then study the impact of combining flavor-complementary detectors on the distance measurement precision. Figure~\ref{fig:CLdistance_SM} shows the confidence bands for a $9M_\odot$ progenitor in the SM, under both the NMO and the IMO, for both individual experiments and detector pairs. Similarly to~\cite{segerlund}, we find that the IMO is associated with the smallest uncertainties. When considering detector pairs, the largest precision increase on the CCSN distance is observed in the IMO for the HK+IceCube combination. However, combining flavor-complementary experiments does not lead to a significant precision increase.

To study the impact of neutrino decays on CCSN distance measurements we consider a typical measurement for a $11~M_\odot$ progenitor and for an example neutrino decay scenario with $\bar r=5$ and $\zeta=1$, under which a large fraction of heavy neutrinos and antineutrinos will have decayed before reaching Earth. This model is associated with a sizable enhancement (reduction) of the $\nu_e$ and $\bar{\nu}_e$ rates in the NMO (IMO), and with a reduction of the total rate expected at DarkSide-20k or ARGO. As discussed in Section~\ref{subsec:flavor}, the case of neutrino decays in the IMO illustrates particularly well the range of possible outcomes for new physics models, as a wide range of new physics scenarios, such as active-sterile neutrino conversions, are also associated with a flavor-dependent flux suppression~\cite{sterile,sterile2,spinflip}. In what follows, we hence focus on neutrino decay scenarios occurring in the IMO. The NMO case, which, for the time windows considered, leads to significantly milder flux distortions compared to the SM, is discussed in Appendix~\ref{appendix:distance_NMO}. In this scenario, when considering neutrino decays, the size of the $90$\% C.I. on the CCSN distance can be reduced by up to $60$\% by combining HK or IceCube with either DUNE or ARGO. The superior performance of the HK+ARGO and IceCube+ARGO pairs in the NMO can be explained by the opposite evolution of the total CCSN rates with $\bar r$ in WC and CE$\nu$NS detectors, which allow breaking the degeneracy between $\bar r$ and the CCSN distance. These results therefore highlight the synergy potential of the three types of detectors considered in this paper.

\begin{figure*}
    \includegraphics[width=\linewidth]{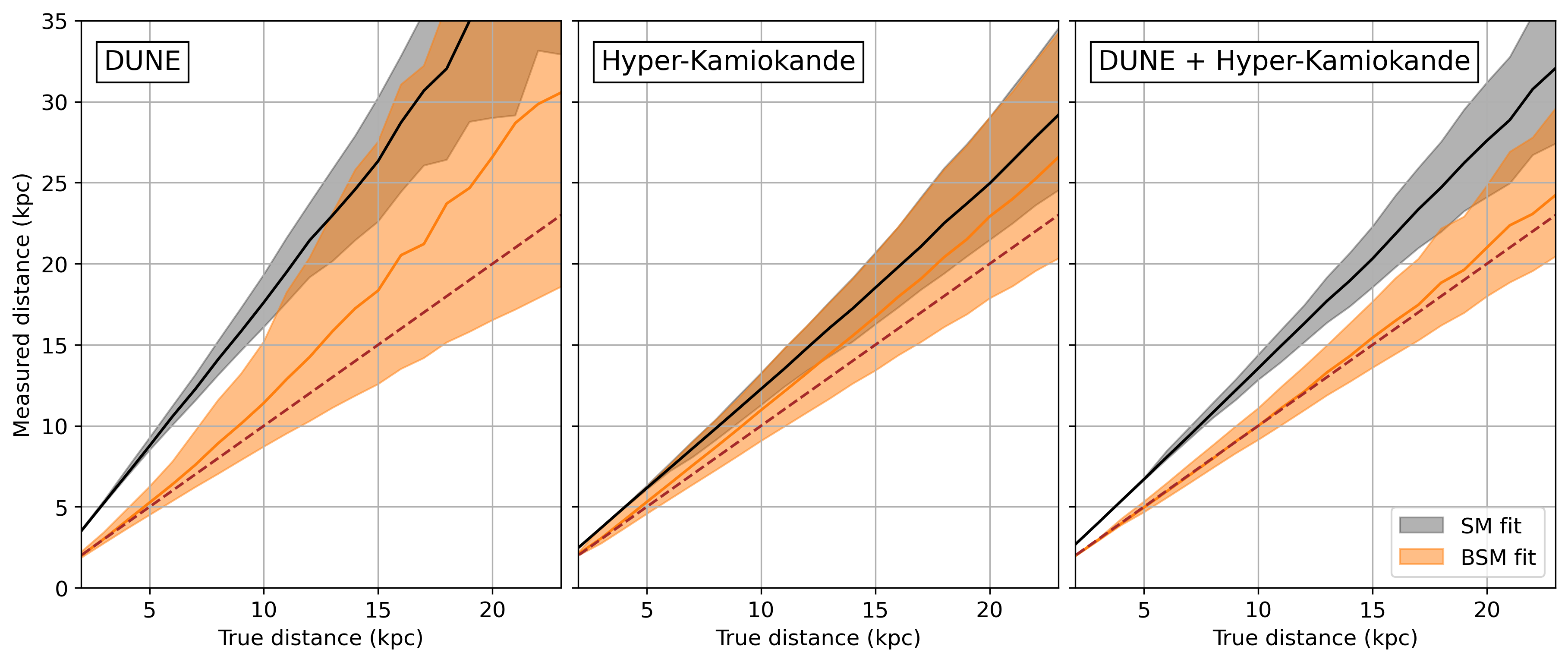}
    \caption{Median values and $90\%$ confidence intervals for the measured CCSN distance, as a function of the true distance for a $11M_\odot$ progenitor and for a neutrino decay model with $\bar{r}=5,\zeta=1$. The IMO is assumed. The grey band shows the results of the fit under the SM hypothesis ($\mathcal{L}_\mathrm{SM}$ in equation~\ref{eq:Ldistance}) and the orange band shows a fit where $\bar{r},\zeta$ are optimized along with the other parameters ($\mathcal{L}_\mathrm{SM}$ in equation~\ref{eq:Ldistance}). The experiments considered are DUNE (left), HK (middle), and DUNE+HK (right). 
    }
    \label{fig:distance_BSM}
\end{figure*}

\begin{figure*}
    \centering    \includegraphics[width=0.6\linewidth]{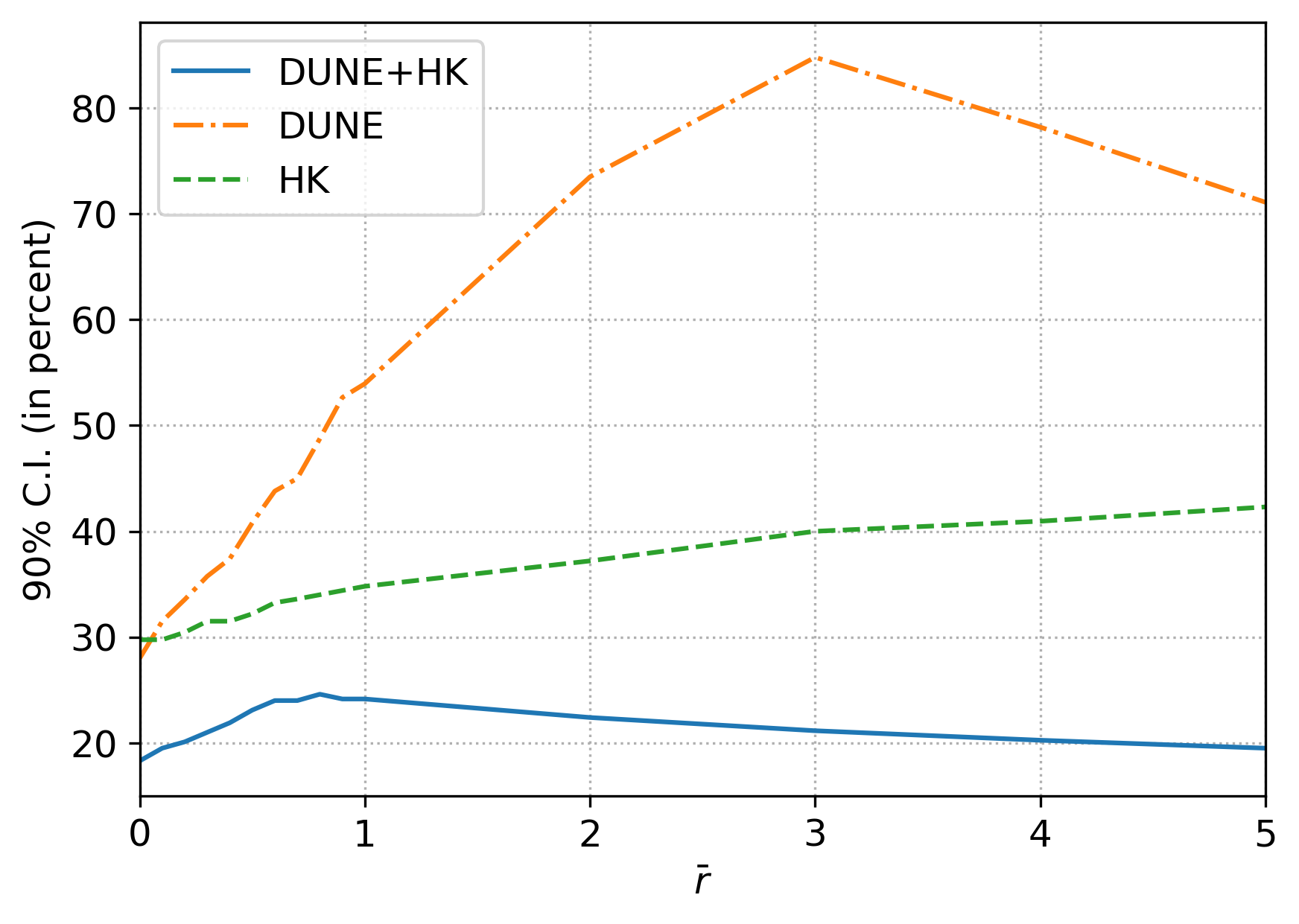}
    \caption{Size of the $90\%$ confidence interval on the measured CCSN distance as a function of $\bar{r}$ for $\zeta=1$, normalized by the true distance, for a supernova at 10~kpc with a $11~M_\odot$ progenitor. The experiments shown here are DUNE (dash-dotted orange), Hyper-Kamiokande (dashed green), and the combination of the two detectors (solid blue).}
    \label{fig:CLdistance_rbar}
\end{figure*}

Figure~\ref{fig:distance_BSM} shows the measured CCSN distance measured and the $90\%$ confidence interval as a function of the true distance for the IMO $(\bar{r}=5,\zeta=1)$ scenario described above, for the DUNE and HK detectors as well as for their combination. In this Figure, the distance estimated using the SM-only likelihood is compared to the distance obtained by allowing $\bar{r}$ and $\zeta$ to vary. 
For DUNE and HK, a clear bias in the distance measurement is observed under the SM assumption. This bias can be at least partially corrected by including $\bar r$ and $\zeta$ in the CCSN distance fit. However, for individual detectors, the addition of these extra degrees of freedom leads to factors of $2$ to $3$ increases of the uncertainties compared to the SM measurements shown in Figure~\ref{fig:distance_1detector}. This uncertainty, however, can be significantly reduced by combining DUNE and HK. To assess the universality of this behavior, we study how the size of the $90\%$ C.I. varies with $\bar{r}$ for DUNE, HK, and DUNE+HK. Here, we keep $\zeta=1$ as, in the IMO, the influence of this parameter on neutrino rates is expected to be negligible for both DUNE and HK. The dependence of the CCSN distance measurement precision in $\bar{r}$ is shown in Figure~\ref{fig:CLdistance_rbar}. For HK, we observe a continuous increase in the size of the confidence interval as $\bar{r}$ grows. This increase can be attributed to the fact that the neutrino decays under consideration primarily affect the normalization of the time-dependent rate observed at HK, without significantly altering its shape, as can be seen in Figure~\ref{fig:decay_snrates}.
In the case of DUNE, where decay-induced flux distortions are more pronounced than at HK but the total event count is lower, the uncertainty size initially grows more rapidly compared to HK before decreasing for large $\bar{r}$ values. Interestingly, when we combine data from both DUNE and HK, we maintain distance measurement uncertainties below $25\%$. This combination results in a substantial improvement over using HK alone, with the size of the confidence interval shrinking by $40$\% to $100$\%. This notable behavior underscores how the synergy between these two experiments enables our analysis to capture the most significant characteristics of neutrino decay models.

\begin{figure*}
    \centering    \includegraphics[width=\linewidth]{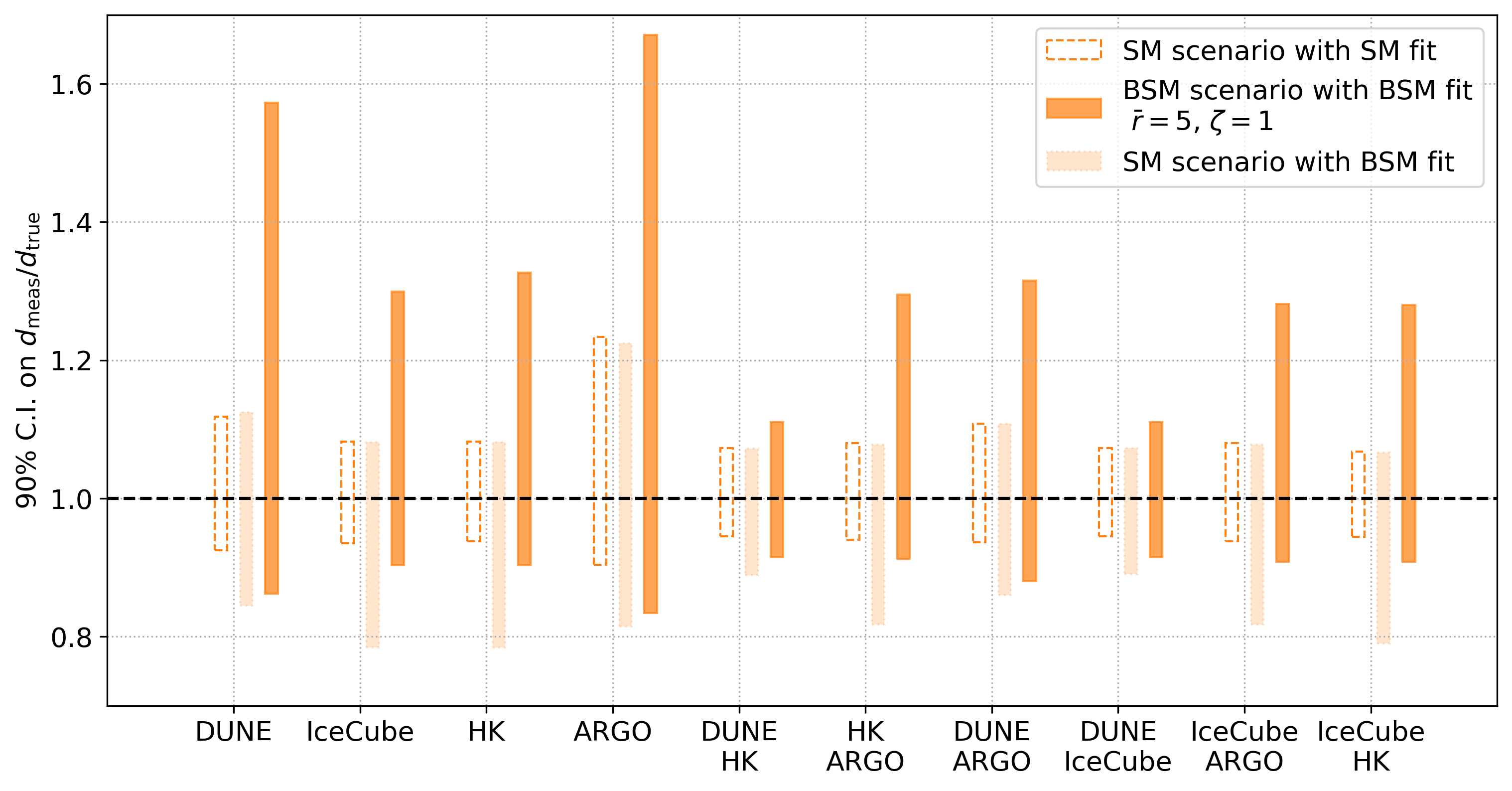}
    \caption{$90\%$ confidence intervals on the measured CCSN distance, normalized by the true distance, for a supernova at 10~kpc, for a $11M_\odot$ progenitor and different detector combinations. The hollow rectangles with a dashed border show the confidence intervals for the SM under the SM hypothesis. The light and dark orange rectangles show the confidence intervals for the SM and for a neutrino decay model with $\bar{r}=5,\zeta=1$, respectively, under the BSM assumption: both measurements are fitted by optimizing the CCSN progenitor mass and location, and the $\bar{r},\zeta$ parameters. The IMO is assumed in both situations. Different detectors and pairs of detectors are considered.}
    \label{fig:CLdistance_BSM}
\end{figure*}

To investigate variations of the CCSN distance precision with the choice of experiment, we compute the $90\%$ C.I. on the CCSN distance for the same experiments as the ones considered in Figure~\ref{fig:CLdistance_SM} for the SM. In Figure~\ref{fig:CLdistance_BSM}, we compare these confidence intervals for a SM measurement, both under the SM and BSM assumptions. We also revisit the scenario discussed earlier, that is, neutrino decays with $\bar{r}=5$ and $\zeta=1$, under the BSM assumption. For each individual detector, we observe a consistent trend: the uncertainties grow as $\bar{r}$ increases. This trend aligns with our previous observations in the DUNE and HK cases. As previously discussed, combining detectors that are sensitive to different neutrino flavors proves effective in mitigating this uncertainty growth. Specifically, when $\bar{r}=5$, we see that combining HK or IceCube with DUNE reduces uncertainties by a factor of two. Finally, note that the introduction of uniform priors on $\bar{r}$ and $\zeta$ for the BSM likelihood causes the confidence intervals to be tilted towards small distances, particularly towards smaller distances, for HK and IceCube. 
Encouragingly, this bias towards small distances can be mitigated by pairing either detector with DUNE or ARGO.

While the confidence limits discussed above have been obtained assuming the neutrino mass ordering is known, we verified that relaxing this assumption does not change our conclusions. Specifically, when treating the mass ordering as another nuisance parameter, the precision of the distance measurements does not change when considering Galactic CCSNe and SM neutrino interactions. When neutrino decays are introduced, not knowing the mass ordering leads to a significant loss of precision for DUNE but the confidence intervals for the other experiments remain stable. Indeed, as discussed in Section~\ref{sec:nmo_full}, HK and IceCube can discriminate between the IMO and NMO scenarios well beyond the Galactic Center, while DUNE can distinguish the two mass orderings only in the SM.

In the context of neutrino decays, we demonstrated that the presence of new physics in the neutrino sector can significantly bias CCSN distance estimates, even for models allowed by current neutrino observations. In particular, new physics scenarios leading to a neutrino flux suppression, such as neutrino decays in the IMO but also active-sterile neutrino conversions~\cite{sterile,sterile2,spinflip}, can lead to overestimating the CCSN distance by up to $70\%$ for a CCSN in the Galactic bulge. Such an estimate would place the CCSN in a region obscured by dust, accessible only by telescopes with a deep but narrow field of view like the Vera Rubin Observatory~\cite{distance_dust}. This result underscores the necessity for all types of telescopes to investigate CCSN alerts, even when they initially appear beyond their observational reach. Additionally, when we introduce new physics parameters for combinations like DUNE+HK and DUNE+IceCube, we observe limited growth in uncertainties. This suggests that with the next generation of neutrino experiments, SNEWS could contemplate incorporating a range of BSM scenarios into its CCSN parameter fits. Importantly, our algorithm's efficiency, with runtimes of under $10$ seconds for detector pairs in a \texttt{Jupyter} notebook, allows for parallel testing of multiple models without compromising the alert system's performance. Finally, we note that this testing procedure would greatly benefit from preliminary analyses capable of identifying and characterizing deviations from the SM before conducting the distance fits. We present examples of such analyses in the next section.


\section{Neutrino decay characterization}
\label{sec:decay_full}
We now describe how to constrain neutrino decay parameters using the likelihood function introduced in equation~\ref{eq:likelihood_sn}. Since neutrino decay signatures can be similar to SM scenarios with a wrong mass ordering or CCSN distance~\cite{degouvea}, we propose algorithms to either identify deviations from the SM without making assumptions about the possible BSM processes, or directly constrain new physics parameters when considering a specific model. Here again, we assume that the neutrino mass ordering is known.
\subsection{Identifying beyond-the-SM phenomena}
\label{subsubsec:bsm_indep}
Since accounting for new physics scenarios when locating CCSNe can be a computationally intensive procedure, we propose to first search for the presence of neutrino two-body decays using hypothesis testing. For a given set of measurements $\{\mathcal{O}_\mathrm{obs}\}$, we take the optimized SM likelihood $\mathcal{L}_\mathrm{SM}$ as a test statistic:
\begin{align}
    \mathcal{L}_\mathrm{SM} = \mathrm{max}_{d,M}[\mathcal{L}(\{\mathcal{O}_\mathrm{obs}\}|d, M, \bar r= 0,\zeta=1)]
    \label{eq:teststat_like}
\end{align}
Similarly to the procedure described in Section~\ref{sec:nmo_full}, we generate pseudo-experiments under the SM hypothesis to evaluate the probability distribution of $\mathcal{L}_\mathrm{SM}$. Here again, we randomly select CCSN models from~\cite{model} with probability $w(M)$. Since we expect this search to be sensitive only to CCSNe in the Milky Way, we only consider CCSN distances in the $0.5-25$~kpc range. For each set of decay parameters $(\bar r,\zeta)$, we then compute the p-value associated with the typical measurement defined in Section~\ref{subsec:likelihood}. 

\begin{figure*}
    \includegraphics[width=\linewidth]{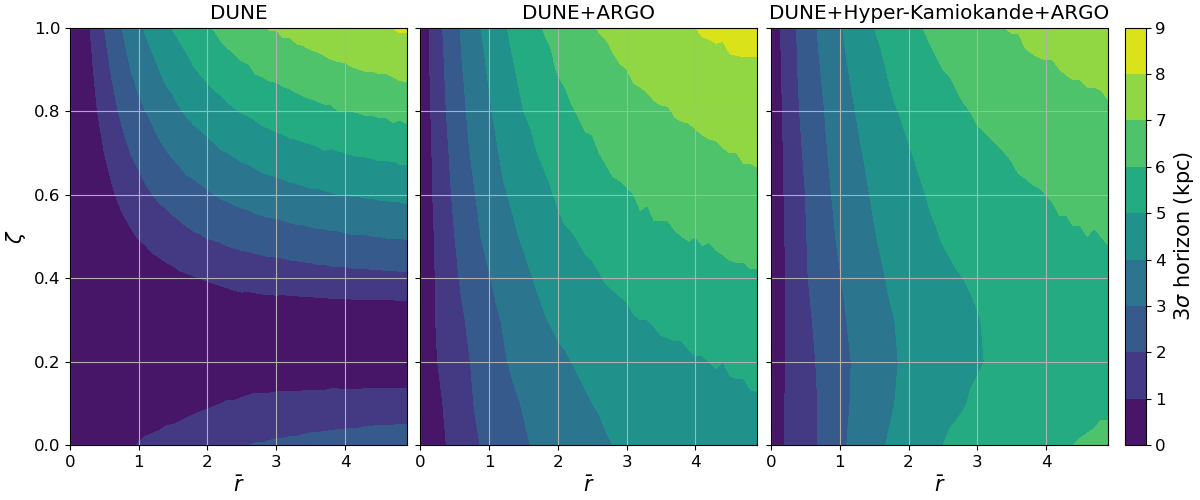}
    \caption{Distance at which a ``typical`` observation would correspond to a $3\sigma$ deviation from the SM, using the test statistic defined in equation~\ref{eq:teststat_like}, as a function of $\bar{r},\zeta$ for a $11M_\odot$ progenitor under the NMO. The distance horizons are shown for DUNE alone (left), DUNE+ARGO (middle), and DUNE+ARGO+HK (right).}
    \label{fig:SMtest_nmo}
\end{figure*}

\begin{figure*}
    \includegraphics[width=\linewidth]{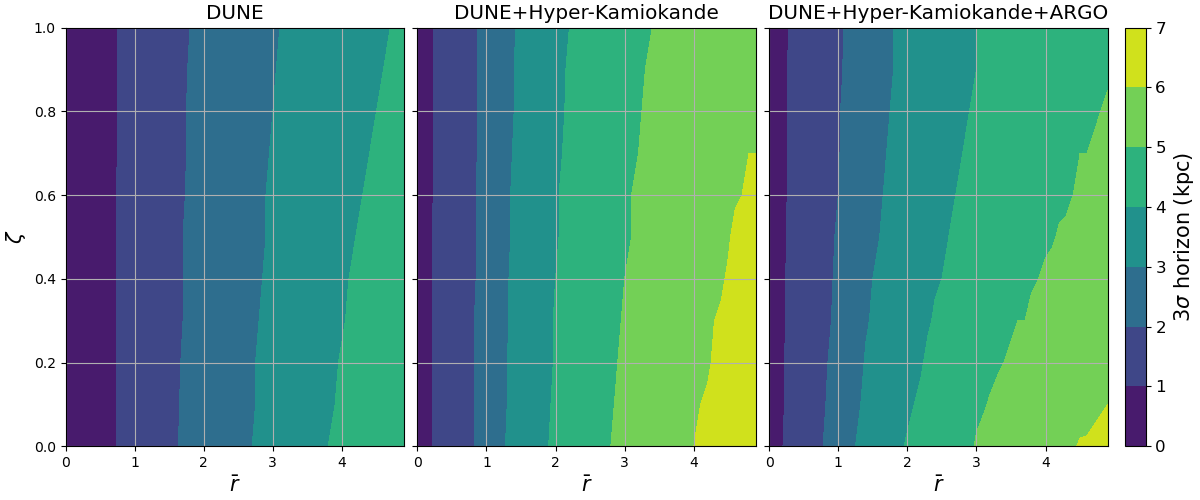}
    \caption{Distance at which a ``typical`` observation would correspond to a $3\sigma$ deviation from the SM, using the test statistic defined in equation~\ref{eq:teststat_like}, as a function of $\bar{r},\zeta$ for a $11M_\odot$ progenitor under the IMO. The distance horizons are shown for DUNE alone (left), DUNE+ARGO (middle), and DUNE+ARGO+HK (right).}
    \label{fig:SMtest_imo}
\end{figure*}

Figures~\ref{fig:SMtest_nmo} and ~\ref{fig:SMtest_imo} show the CCSN distance corresponding to a $3\sigma$ exclusion of the SM as a function of $\bar r$ and $\zeta$, in the NMO and the IMO respectively, for different combinations of measurements at DUNE, HK, and ARGO. Here, we showcase the performance of our approach and the impact of combining experiments by selecting the detector (or set of detectors) with the highest reach over most of the parameter space for each combination of $1$, $2$, and $3$ experiments. For both mass orderings, DUNE is the individual detector associated with the largest $3\sigma$ distance horizons over a large fraction of the $(\bar{r},\zeta)$ space. In the NMO, in particular, models with large $\bar{r}$ and $\zeta$ can be distinguished from the SM due to the partial reappearance of the neutronization peak, as discussed in Section~\ref{subsec:flavor}. For lower values of $\zeta$, however, this neutronization peak is suppressed again, resulting in a near-degeneracy with the SM, particularly around $\zeta = 0.25$. This degeneracy, also present at HK,  can be effectively resolved by combining DUNE with ARGO. ARGO's sensitivity to the sum of all neutrino flavors allows it to discriminate between active neutrino oscillations and decays into sterile neutrinos. Finally, a combined measurement using DUNE, ARGO, and HK, leads to an extended horizon at small $\zeta$ but slightly reduces the horizon at large $\zeta$. For $\bar{r}\gtrsim 3$, distance horizons exceeding $5$~kpc can be achieved regardless of $\zeta$. Conversely, in the IMO, there is no degeneracy region with the SM at large $\bar{r}$ for DUNE as the primary signature of neutrino decays is the suppression of the neutronization peak, which affects different time windows unequally. While combining DUNE with Hyper-Kamiokande increases the distance horizon for all $\zeta$ values,  the inclusion of ARGO results in a decreased horizon, especially for large $\zeta$. Nonetheless, the final reach with the 3-detector combination remains larger than that of DUNE alone for all $\zeta$ values.

The $(\bar{r},\zeta)$ dependence of the $3\sigma$ horizons shown in Figures~\ref{fig:SMtest_imo} and \ref{fig:SMtest_nmo}, and the dependency of these horizons in the types of detectors considered, reveal several noteworthy aspects. First, combining multiple detectors can sometimes lead to a reduction in the search's reach. Indeed, in our analysis, we exclusively examine deviations from the SM rather than fitting neutrino decay parameters. Assuming that we lack prior knowledge of the CCSN progenitor and its location, adding a new detector could potentially make observations more similar to an SM scenario. However, note that this reduction in performance remains limited for the models we have considered. Second, while combining experiments does not always result in a performance boost, it allows to identify new physics scenarios which, for individual experiments, could be completely degenerate with the SM. By combining DUNE, a WC detector, and ARGO, we ensure that we can identify all flavor-dependent distortions in the neutrino flux if the event counts are high enough. Third, the method we propose here stands out for its speed and simplicity, taking less than a second to execute. It provides a single easily interpretable output: the p-value of the measurement within the SM framework. This metric can be assessed not only by organizations like SNEWS but also by individual experiments. Lastly, although our approach is straightforward, rapid, and does not rely on assumptions about the possible BSM processes, it is most suitable for nearby CCSNe. While the choice of a $3\sigma$ criterion can be adjusted based on the requirements of alert systems and telescopes, exploring regions like the Galactic Center and beyond may necessitate the consideration of specific BSM scenarios.


\subsection{Fitting neutrino decay parameters}
In addition to searching for deviations from the SM without assuming a specific new physics scenario, we investigate the impact of combining flavor-complementary detectors when fitting the parameters of a given BSM scenario, again focusing on neutrino two-body decays. As discussed above, this model-dependent approach could allow identifying deviations from for the SM for CCSNe beyond a few kiloparsecs. To evaluate the $(\bar r,\zeta)$ parameters, we maximize the likelihood from equation~\ref{eq:likelihood_sn} over $d$ and $M$, assuming that the neutrino mass ordering is known. 

\begin{figure*}
\includegraphics[width=\linewidth]{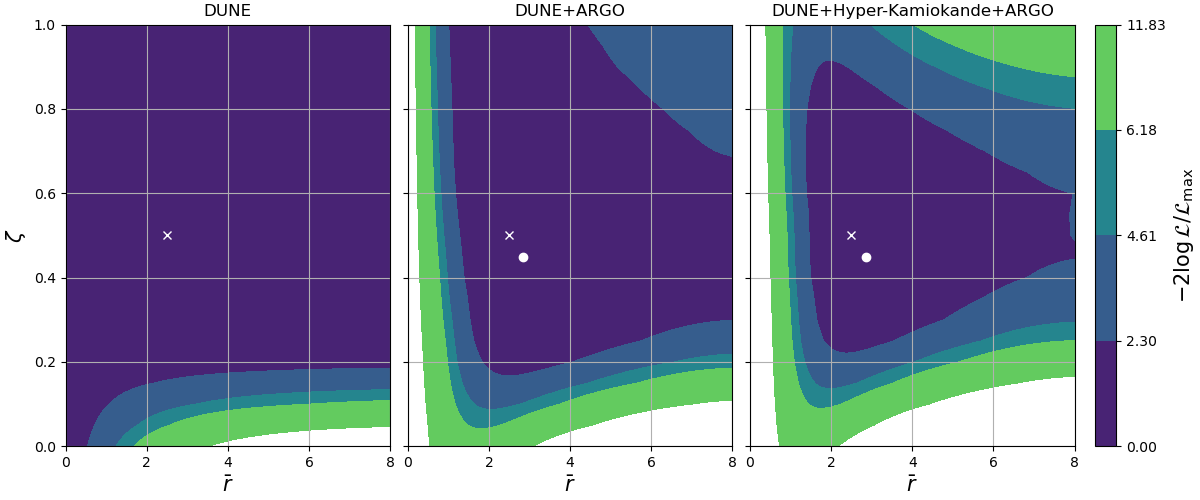}
    \caption{Best-fit point and isolikelihood contours corresponding to $\mathcal{L}_\mathrm{max}+2.3, 4.6, 6.2$, and $11.8$ (from light to dark) in the $(\bar r, \zeta)$ space for a $11M_\odot$ progenitor and a neutrino decay scenario with $\bar{r}=2.5,\zeta=0.5$, assuming the NMO. The contours are shown for DUNE alone (left), DUNE+ARGO (middle), and DUNE+ARGO+HK (right). The white dot and white cross show the best-fit and the true values of $\bar{r}$ and $\zeta$, respectively.}
    \label{fig:decay_constraints_BSM_nmo}
\end{figure*}

\begin{figure*}
\includegraphics[width=\linewidth]{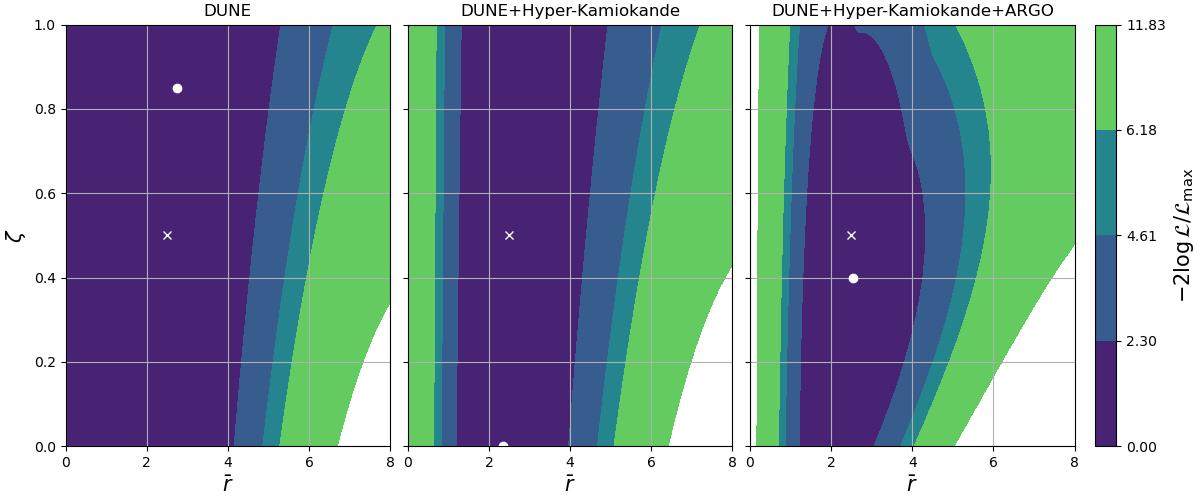}
    \caption{Best-fit point and isolikelihood contours corresponding to $\mathcal{L}_\mathrm{max}+2.3, 4.6, 6.2$, and $11.8$ (from light to dark) in the $(\bar r, \zeta)$ space for a $9M_\odot$ progenitor and a neutrino decay scenario with $\bar{r}=2.5,\zeta=0.5$, assuming the IMO. The contours are shown for DUNE alone (left), DUNE+HK (middle), and DUNE+HK+ARGO (right).The white dot and white cross show the best-fit and the true values of $\bar{r}$ and $\zeta$, respectively.}
    \label{fig:decay_constraints_BSM_imo}
\end{figure*}

To assess the capability of our approach to identify BSM scenarios, we consider a model with $\bar{r}=2.5$ and $\zeta=0.5$. With the generic approach described in Section~\ref{subsubsec:bsm_indep}, this model could be distinguished from the SM at $3\sigma$ only for CCSNe closer than $4-5$~kpc. Here, similarly to the SM scenario discussed above, we show the best-fit point and the $-2\log\mathcal{L}_\mathrm{max}+2.3$, $4.6$, $6.2$, and $11.8$ contours (which would have corresponded to the $1\sigma$, $90\%$ confidence level, $2\sigma$ and $3\sigma$ values if Wilks' theorem were applicable) for the NMO and the IMO in Figures~\ref{fig:decay_constraints_BSM_nmo} and~\ref{fig:decay_constraints_BSM_imo}, respectively, for a $11~M_\odot$ progenitor.  
Here again, combining flavor-complementary experiments significantly increases the precision on $\bar r$ and $\zeta$ and breaks the degeneracies observed for DUNE and HK in the NMO. For both mass orderings, including ARGO sizeably increases the sensitivity in $\zeta$. Moreover, for the DUNE+HK+ARGO combination, the likelihood for $\bar r=0$ lies outside the $-2\log\mathcal{L}_\mathrm{max}+11.8$ contour, revealing a considerable performance increase compared to the generic approach from Section~\ref{subsubsec:bsm_indep}. Hence, for a wide range of neutrino decay models, evidence for BSM phenomena could be detected for CCSNe as far as the Galactic bulge. Finally, note that one possible caveat of this approach is its runtime: in \texttt{Python}, generating the contours shown in Figures~\ref{fig:decay_constraints_BSM_nmo} and ~\ref{fig:decay_constraints_BSM_imo} for a combination of $3$ detectors takes $\sim100$~s. However, for these Figures, a particularly fine $(\bar{r},\zeta)$ grid was used, with steps of $0.05$ for both parameters. Simply dividing the size of this grid by $4$ would bring the runtime of this algorithm under one minute with minimal loss of information.


\section{Discussion}
\label{sec:discussion}
With a minimal set of observables, we have demonstrated that harnessing the capabilities of the next generation of neutrino experiments can enable CCSN alert systems to derive meaningful constraints on CCSN properties without the need for event-by-event energy reconstruction.
Taking neutrino decays as an example, we have illustrated how the presence of new physics in the neutrino sector can introduce significant biases into CCSN distance estimates. Such biases could have consequences for electromagnetic follow-ups. In particular, if the source is estimated to be in the foreground of the Galactic Center, optical telescopes could be the first to try to identify it, but might fail to do so because of the significant absorption on the line of sight (see e.g.~\cite{distance_dust}). However, we have also shown that when new physics affects neutrino fluxes in a flavor-dependent manner, it is possible to correct distance estimates while keeping uncertainties under control. This correction can be achieved by combining data from WC detectors like HK and IceCube with data from DUNE.
Furthermore, we have highlighted that deviations from the SM can be efficiently characterized by combining information from WC detectors, DUNE, and ARGO. While a generic search for anomalies would only detect these deviations for CCSNe within a few kiloparsecs, tailored fits targeting new physics models could potentially extend our reach to CCSNe in the Galactic bulge.
The findings presented here also point to several promising avenues for future exploration, aimed at optimizing realtime CCSN identification.

\paragraph{CCSN model dependence of early neutrino rates} 
Because of the need to compute a large number of progenitors, the supernova modelling used for our study is restricted to a one-dimensional treatment and is therefore approximate in several aspects. Multi-dimensional hydrodynamic instabilities such as PNS convection, neutrino-driven convection and the Standing Accretion Shock Instability (SASI)~\cite{muller2020} are missing from this modelling. 
Note, however, that the instabilities do not grow immediately after bounce: PNS convection starts $\sim 50$ ms after bounce~\cite{Nagakura2020}, prompt post-shock convection starts $\sim50-100$ ms after bounce~\cite{muller2020} and SASI reaches significant amplitudes at still later times of $\mathcal{O}(100~\mathrm{ms})$~\cite{muller2020,foglizzo2015}.  As a consequence, the early-time dynamics and neutrino rates during the neutronization phase should not be affected by the restriction to one dimension. On the other hand, the later accretion could be impacted by the multidimensional dynamics, and its impact on our results would deserve to be addressed in future studies. \\
Another important assumption of the supernova modelling is the neglect of rotation and magnetic fields, which are not expected to play a dominant role in the explosion dynamics for most stellar progenitors \cite{janka2012}.
However, it has been shown that neutrino-driven post-shock convection~\cite{endeve2012} and moderate rotation rates \cite{mullerAndvarma2020} can amplify weak magnetic field seeds up to the point where they can assist the neutrino-heating mechanism and quantitatively affect the supernova explosion.
Since these results have to rely on computationally expensive three-dimensional CCSN models, more studies are still needed in order to accurately assess the impact of rotation and magnetic fields on the neutrino emission from stellar explosions.

\paragraph{Parameterization of new physics effects} As discussed in Section~\ref{subsec:flavor}, new physics scenarios with the potential to impact early CCSN neutrino rates can be currently classified into three broad categories. Each of these categories has been studied using simplified frameworks involving only two or three new parameters~\cite{degouvea,self-interactions,sterile,spinflip,ando_decay,volpe_decay}. Consequently, incorporating the effects of new physics into realtime CCSN analyses necessitates the establishment of a strategy for exploring these diverse frameworks. One approach could involve an alert system simultaneously fitting several representative BSM scenarios. However, considering the limited information employed by our algorithm, as well as the current approaches for CCSN distance measurements~\cite{segerlund}, it is likely that multiple new physics models will yield similar predictions. Therefore, the development of a generic parameterization capable of encompassing multiple classes of BSM scenarios would greatly enhance the analytical capabilities of alert systems. It is also worth noting that our analysis exclusively considered flat prior probabilities for new physics parameters. While transitioning to other realistic prior distributions is unlikely to qualitatively alter our results, determining the optimal prior choices at alert systems, in collaboration with theorists, will be essential for implementing the final analysis.

\paragraph{Neutrino energies} While the approach discussed here used only the timing information of neutrino events, it could be extended to include neutrino energy information. The inclusion of this energy information, however, would need to be done on an experiment-by-experiment basis since detectors such as IceCube, KM3NeT, DarkSide-20k, and ARGO, could only constrain global properties of CCSN neutrino spectra. 

\paragraph{Large-scale CE$\nu$NS detectors} In this study, we have demonstrated the potential of experiments sensitive to the sum of all neutrino flavors (through CE$\nu$NS) for CCSN studies, focusing on DarkSide-20k and ARGO. While DarkSide-20k may lack the capacity to constrain CCSN and neutrino properties effectively, ARGO, when combined with other experiments, could significantly enhance the sensitivity to mass ordering and the investigation of active neutrino decays into sterile neutrinos. It is important to note, however, that ARGO is currently in its early planning stages, and its commissioning date remains uncertain.
An alternative option worth considering is the JUNO experiment, which possesses sensitivity to both $\bar{\nu}_e$ and CE$\nu$NS interactions. Specifically, for a CCSN located at the Galactic Center, we anticipate approximately $2000$ CE$\nu$NS interactions at JUNO~\cite{juno,juno2}. Depending on its detection efficiency, JUNO has the potential to achieve a sensitivity level comparable to, or even surpassing, that of ARGO. To facilitate realistic performance forecasts for alert systems, it is essential for the JUNO collaboration to release detailed, energy-dependent detection efficiency curves for CE$\nu$NS. This data would enable accurate assessments of the experiment's capabilities.

\paragraph{A unified strategy at alert systems} 
In this study, we have presented various applications of the analysis method detailed in Section~\ref{sec:likelihood_approach}. For an alert system like SNEWS, consolidating these diverse applications into a cohesive procedure is of utmost importance.
Our proposed approach begins by searching for and characterizing deviations from the SM using the methodologies outlined in Section~\ref{sec:decay_full}.
Furthermore, if measurements allow for the existence of BSM scenarios capable of significantly distorting CCSN neutrino fluxes, we can implement dedicated CCSN distance fits. These fits can follow a similar framework to the one described in Section~\ref{sec:discussion} for neutrino decays. Additionally, the neutrino MO can be determined separately using the approach discussed in Section~\ref{sec:nmo_full}, if this MO has not been already measured by experiments targeting neutrino oscillations. This step, which takes less than a second, would allow verifying the validity of the constraints on BSM neutrino interactions derived earlier and would yield a key result for a wide variety of neutrino searches. The layered approach proposed here is facilitated by the efficiency of the algorithms involved, with each step requiring less than a minute of computation using the current \texttt{Python} implementation.\newline\\

By conducting a straightforward likelihood analysis, our research underscores the capability of upcoming experiments to bolster the effectiveness of CCSN alert systems. Our study notably evaluates the impact of new physics on CCSN distance estimations and highlights the importance of combining data from various detectors in order to simultaneously constrain supernova properties and the physics of the neutrino sector. Consequently, our research sets the stage for a more accurate realtime CCSN identification process.

\section*{Acknowledgments}
This work is supported by LabEx UnivEarthS (ANR-10-LABX-0023 and ANR-18-IDEX-0001) and Paris Region (DIM ORIGINES). We thank Gwenhaël de Wasseige, Maria Cristina Volpe and Joachim Kopp for interesting and helpful discussions. 
This project has received funding from the European Union's Horizon Europe research and innovation programme under the Marie Sk\l{}odowska-Curie grant agreement No 101064953 (GR-PLUTO). Joao Coelho, Alexis Coleiro, Sonia El Hedri, Davide Franco, Isabel Goos and Antoine Kouchner are supported by Centre National de la Recherche Scientifique (CNRS). The Moroccan Ministry of Higher Education, Scientific Research and Innovation is acknowledged.

\appendix
\section{Impact of neutrino two-body decays on detected rates}
\label{appendix:decays}
We follow the approach described in~\cite{degouvea} and consider only decays of the heaviest (active) neutrino species $\nu_h$ into the lightest (active or sterile) $\nu_\ell$. We consider the following processes:
\begin{align}
    \nu_{h,L} &\rightarrow \nu_{\ell,L/R} + \phi\\
    \nu_{h,L} &\rightarrow \bar{\nu}_{\ell,L/R} + \phi
\end{align}
where $\phi$ is a scalar field and the helicities of the final-state neutrinos and antineutrinos depend on the model considered. In the Dirac case, if the neutrinos (antineutrinos) are left-handed (right-handed), they will be visible in neutrino experiments. Otherwise, they will be sterile and will escape detection. In the case of Majorana neutrinos all final-state neutrinos will be visible.

In the analysis shown in this paper, we consider the Dirac $\phi_0$ model introduced in~\cite{degouvea}, where the neutrinos are Dirac and lepton number is conserved. In this case, a heavy active (anti)neutrino can decay to either another active (anti)neutrino via a \emph{helicity-conserving} interaction or to a sterile (anti)neutrino via a \emph{helicity-flipping} interaction. While the products of the latter will not be detectable at current and upcoming experiments, helicity-conserving decays lead to an increase in the flux of light active neutrinos. The fluxes of the heaviest and lightest active neutrinos as a function of the distance travelled can be expressed as~\cite{degouvea,ando_decay,volpe_decay}:
\begin{align}
    \Phi_h(\bar r, E_\nu,t) =& e^{-\bar r \frac{E_0}{E_\nu}} \Phi_h^{(0)}(E_\nu,t)\\
    \nonumber
    \Phi_\ell(\bar r, E_\nu,t) =& \zeta \int_{E_\nu}^{\infty} \left(1 - e^{-\bar r \frac{E_0}{E'}}\right)\psi(E',E_\nu)\Phi_h^{(0)}(E',t)\,\mathrm{d}E' \\
    \nonumber
    &+ \Phi_\ell^{(0)}(E_\nu,t)
    \label{eq:nudecay_init}
\end{align}
where $\zeta$ is the branching ratio associated with the helicity-conserving decays of $\nu_\ell$ and $\Phi^{(0)}$ represents the neutrino flux right outside the star. Neglecting the masses of $\nu_\ell$ and $\phi$, the kinematic factor $\psi(E_h,E_\ell)$ associated with helicity-conserving decays is:
\begin{align}
    \psi(E_h,E_\ell) = \frac{2E_\ell}{E_h^2}.
\end{align}
Injecting the expression for $\psi$ into  equation~\ref{eq:nudecay_init} and performing the variable change $y = \dfrac{E'}{E_\nu}$, the lightest neutrino flux can be rewritten as
\begin{align}
    \Phi_\ell(\bar r, E_\nu,t) =& 2\zeta \int_{1}^{\infty} \frac{1 - e^{-\bar r \frac{E_0}{yE_\nu}}}{y^2}\Phi_h^{(0)}(y E_\nu,t)\,\mathrm{d}y \\
    \nonumber
    &+ \Phi_\ell^{(0)}(E_\nu,t).
    \label{eq:nudecay2}
\end{align}
When traveling through the star, neutrinos undergo adiabatic MSW flavor transitions which turn electron neutrinos (antineutrinos) into $\nu_h$ ($\nu_\ell$) states and turn muon and tau neutrinos and antineutrinos into the other mass eigenstates. Since the muon and tau emissions during the first $\mathcal{O}(100~\mathrm{ms})$ of the CCSN can safely be assumed to be the same, once outside the star we can model the energy spectra of the different neutrino mass eigenstates using pinched Fermi-Dirac distributions~\cite{fermi_dirac}:
\begin{align}
    \Phi_i(E,t) = \frac{(\alpha+1)^{\alpha+1}}{\langle E\rangle^2\Gamma(\alpha+1)}L(t)\left(\frac{E}{\langle E\rangle}\right)^\alpha e^{-(\alpha+1)\frac{E}{\langle E\rangle}}
\end{align}
where $L(t)$ is the luminosity, $\langle E\rangle$ is the mean energy, and $\alpha$ is a dimensionless parameter. The lightest neutrino flux can therefore be expressed as:
\begin{align}
    \Phi_\ell(\bar r, E_\nu,t) =& 2\zeta \frac{(\alpha+1)^{\alpha+1}}{\langle E\rangle^2\Gamma(\alpha+1)}L(t)\left(\frac{E_\nu}{\langle E\rangle}\right)^\alpha\\
    \nonumber
    &\times\int_{1}^{\infty} \left(1 - e^{-\bar r \frac{E_0}{yE_\nu}}\right) y^{\alpha-2}e^{-(\alpha+1)\frac{y E_\nu}{\langle E\rangle}}\,\mathrm{d}y \\
    \nonumber
    &+ \Phi_\ell^{(0)}(E_\nu,t).
    \label{eq:nudecay3}
\end{align}

\section{Analytical approximation for likelihoods}
\label{appendix:interpolation}
To compute the likelihood shown in equation~\ref{eq:likelihood_sn} we evaluate the detection rates $N_i$ for each detector, for all the models described in Section~\ref{subsec:models}, and for the $(\bar r,\zeta)$ values given by Equation~\ref{eq:decay_grid}.
To interpolate between the values from the grid, we fit the number of neutrino events in each time bin by a fifth-degree polynomial in $\bar r$ whose coefficients depend linearly on $\zeta$:
\begin{align}
    N(\bar r,\zeta) &= \sum_{i=0}^5 C_i(\zeta)\bar r^i\\
    C_i(\zeta) &= A_i \zeta + B_i.
    \label{eq:likepoly}
\end{align}

\begin{figure*}
    \centering    \includegraphics[width=0.49\linewidth]{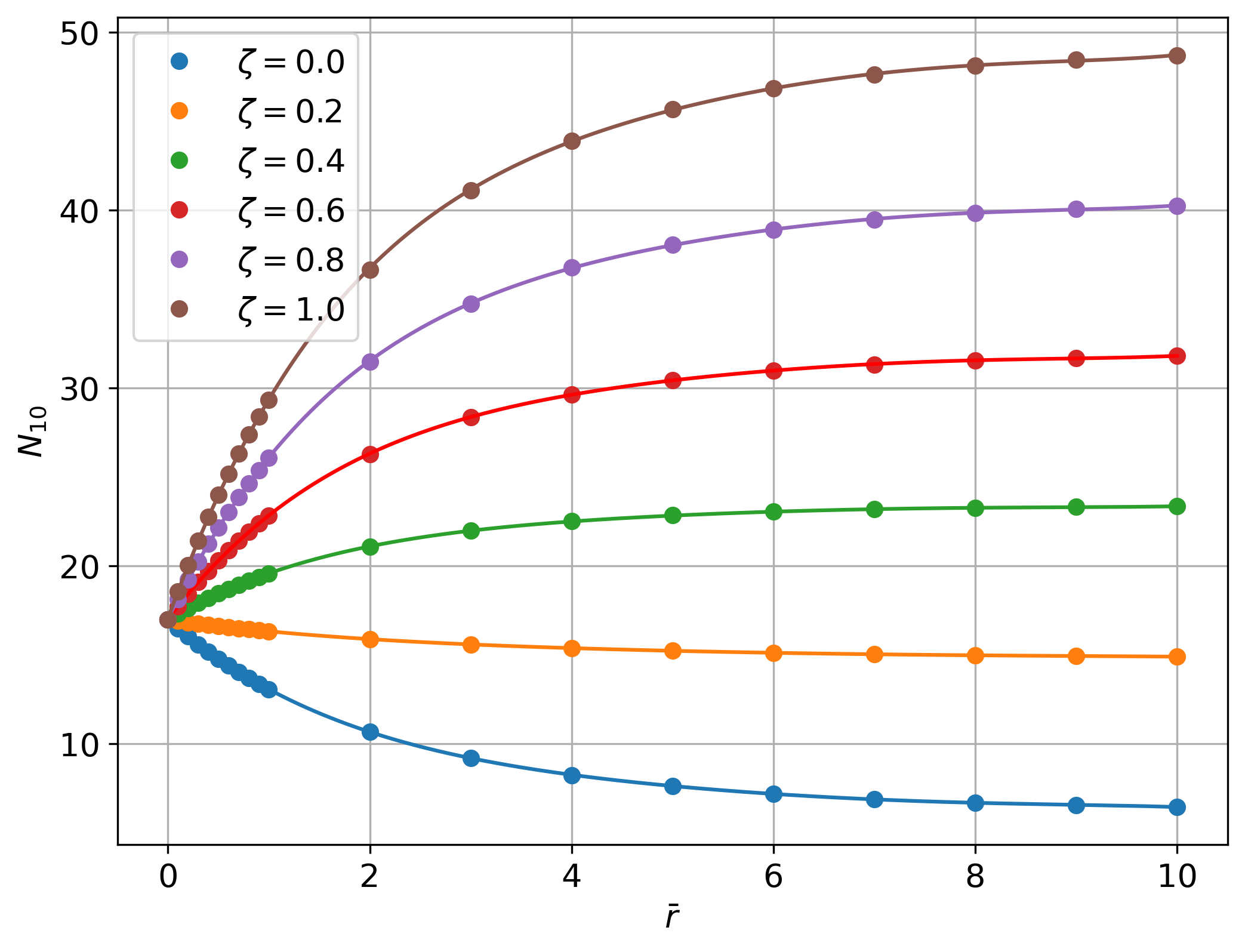}
    \includegraphics[width=0.49\linewidth]{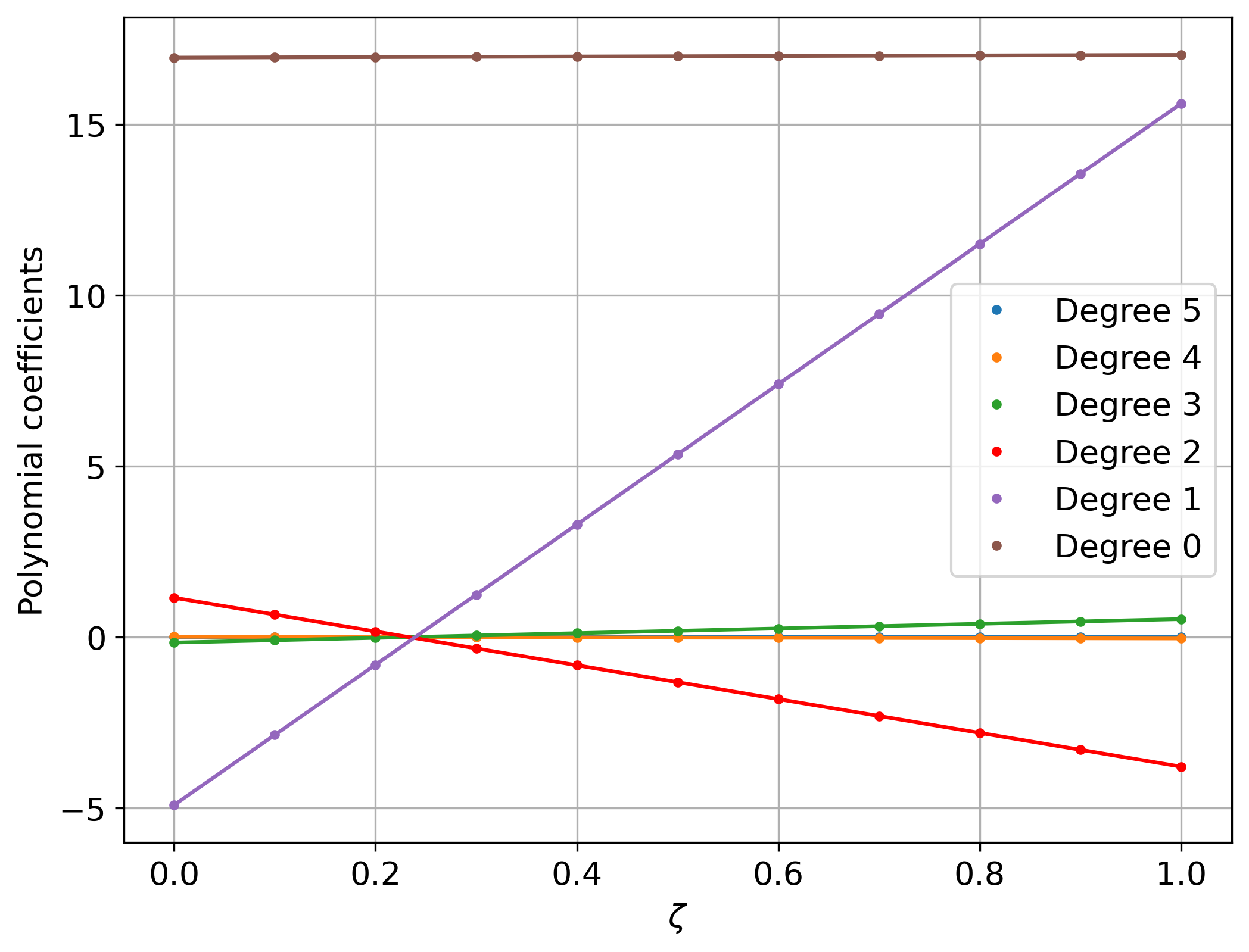}
    \caption{Interpolation of the number of events expected at DUNE in the NMO for a $9~M_\odot$ progenitor model, within the first $10$~ms after the beginning of the CCSN. Left: expected number of events as a function of $\bar r$ with a 5-degree polynomial fit. Right: $C_i$ coefficients of the 5-degree polynomial fit of $\bar r$ as a function of $\zeta$.}
    \label{fig:interp_decay}
\end{figure*}

Each CCSN model, detector, mass ordering, and time window will be associated to a set of polynomial coefficients $(A_i,B_i)$. An example of this polynomial fit is shown in Figure~\ref{fig:interp_decay} for the number of events expected within the first $10$~ms after the detection of a CCSN with a $9~M_\odot$ progenitor at the DUNE detector, in the NMO. In the case of DUNE, discrepancies between the analytical model and the exact rate values are below $1\%$. The largest discrepancies have been observed for event rate predictions at DarkSide-20k and ARGO within the first $10$~ms of the CCSN and remain below $5\%$. We therefore neglect the associated errors in our subsequent analyses.

\section{Likelihood optimization over the supernova distance}
\label{appendix:distance}
To optimize the likelihood over the CCSN distance, we use the fact that the expected number of signal events at a given experiment is proportional to the inverse of the CCSN distance squared. We can therefore rewrite the expectation values $\lambda_i$ in equation~\ref{eq:poisson_like} as:
\begin{align}
    \lambda_i = S_i \left(\frac{d_0}{d}\right)^2 + B_i
\end{align}
where $B_i$ is the expected number of background events and $S_i$ is the expected number of signal events at $d_0 = 10~$kpc. Optimizing the likelihood from equation~\ref{eq:poisson_like} then amounts to solving the following equation:
\begin{equation}
\frac{\mathrm{d}\log\mathcal{L}(\{\mathcal{O}_\mathrm{obs}\} | d, M, \bar r, \zeta, \mathrm{MO})}{\mathrm{d}X} = \sum_{i} S_i\left(\frac{N_i}{S_i X + B_i} - 1\right) = 0
\label{eq:likelihood_derivative}
\end{equation}
where $X = \left(\dfrac{d_0}{d}\right)^2$ and the $i$ index runs over all time bins and detectors considered. If all the backgrounds are negligible, this equation can be solved analytically, giving:
\begin{align}
    X = \frac{\sum_i N_i}{\sum_i S_i}.
    \label{eq:distance_noBG}
\end{align}
If not all backgrounds can be neglected, we solve equation~\ref{eq:likelihood_derivative} numerically. Since the left-hand side of the equation decreases monotonically for $X \geq 0$, we use Newton's method with the following initial guess:
\begin{align}
    X_\mathrm{init} = \mathrm{max}\left\{\frac{\sum_i N_i - B_i}{\sum_i S_i}, X_\mathrm{min}+1\right\}
    \label{eq:dist_guess}
\end{align}
where $X_\mathrm{min}=\left(\frac{d_0}{d_\mathrm{max}}\right)$ and $d_\mathrm{max}$ is the maximal CCSN distance considered for a given analysis (we define $X_\mathrm{min}$ only when we use priors on CCSN distances, in Sections~\ref{sec:nmo_full} and \ref{subsubsec:bsm_indep}). Additionally, the left-hand side of equation~\ref{eq:likelihood_derivative} can diverge for $X\rightarrow 0$. In these cases, if Newton's method converges towards $X \ll \mathrm{min}_i \dfrac{B_i}{S_i}$, for $B_i>0$, we perform a first-order expansion of equation~\ref{eq:likelihood_derivative} in $X$:
\begin{equation}
\frac{\mathrm{d}\log\mathcal{L}(\{\mathcal{O}_\mathrm{obs}\}|d, M, \bar r, \zeta, \mathrm{MO})}{\mathrm{d}X} \approx -X\sum_{i} \frac{N_i S_i^2}{B_i^2} +\left(\sum_{i} S_i\frac{N_i}{B_i} -\sum_i S_i - \sum_j S_j\right) + \frac{\sum_j N_j}{X} = 0
\label{eq:likelihood_derivative_taylor}
\end{equation}
where the indices $i$ and $j$ run over observables at experiments with non-zero and negligible backgrounds, respectively. We then turn this equation into a degree 2 polynomial and solve it analytically. Finally, independently of the distance optimization method, we set
\begin{align}
    X_\mathrm{optimal} = \mathrm{max}\left\{ X_\mathrm{sol}, X_\mathrm{min}\right\}
\end{align}
where $X_\mathrm{sol}$ is the numerical solution found by our algorithm.

\section{CCSN distance measurements: neutrino decays in the NMO}
\label{appendix:distance_NMO}
Section~\ref{sec:discussion} presented CCSN distance measurements in the presence of neutrino decays in the IMO. In this appendix, we present results obtained by applying the same approach for decay scenarios in the NMO.

\begin{figure*}
    \includegraphics[width=\linewidth]{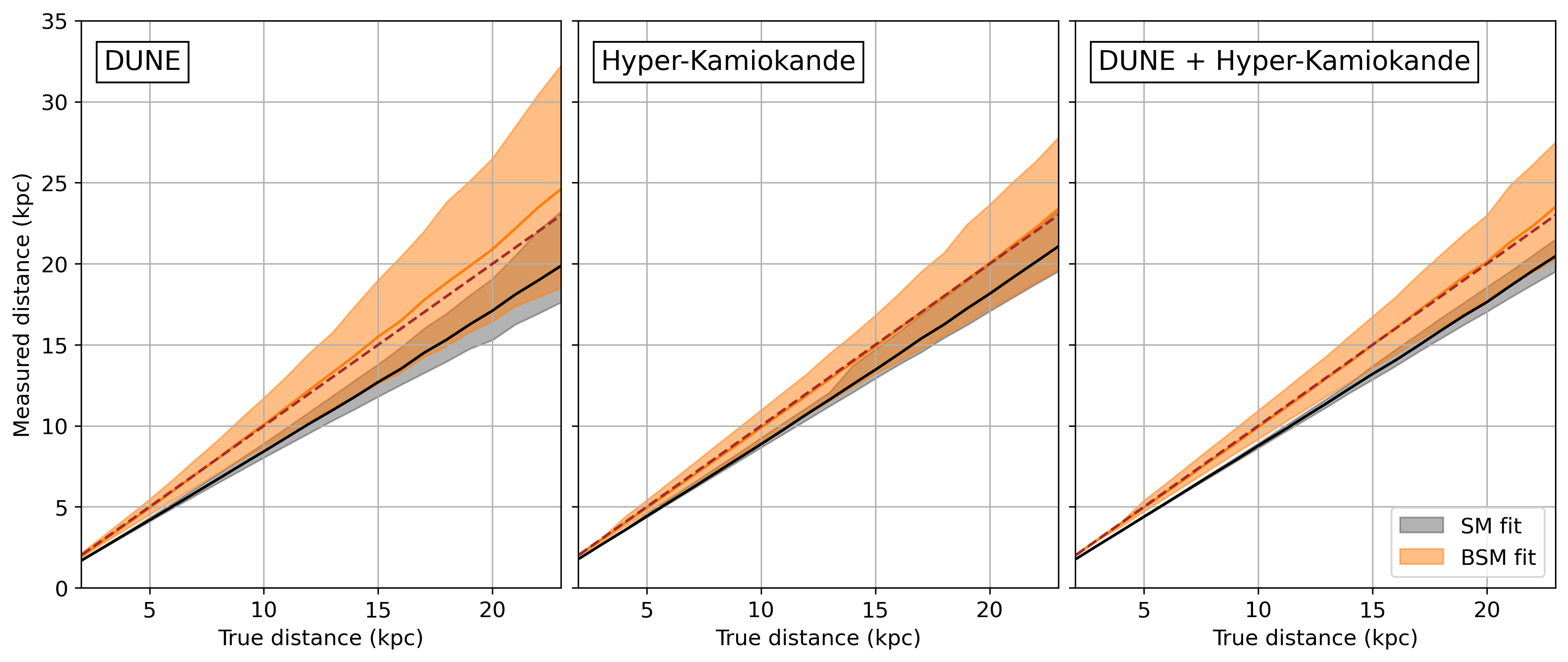}
    \caption{Median values and $90\%$ confidence intervals for the measured CCSN distance, as a function of the true distance for a $11M_\odot$ progenitor and for a neutrino decay model with $\bar{r}=5,\zeta=1$. The NMO is assumed. The grey band shows the results of the fit under the SM hypothesis ($\mathcal{L}_\mathrm{SM}$ in equation~\ref{eq:Ldistance}) and the orange band shows a fit where $\bar{r},\zeta$ are optimized along with the other parameters ($\mathcal{L}_\mathrm{SM}$ in equation~\ref{eq:Ldistance}). The experiments considered are DUNE (left), HK (middle), and DUNE+HK (right). 
    }
    \label{fig:distance_BSM_nmo}
\end{figure*}

\begin{figure*}
    \centering    \includegraphics[width=0.6\linewidth]{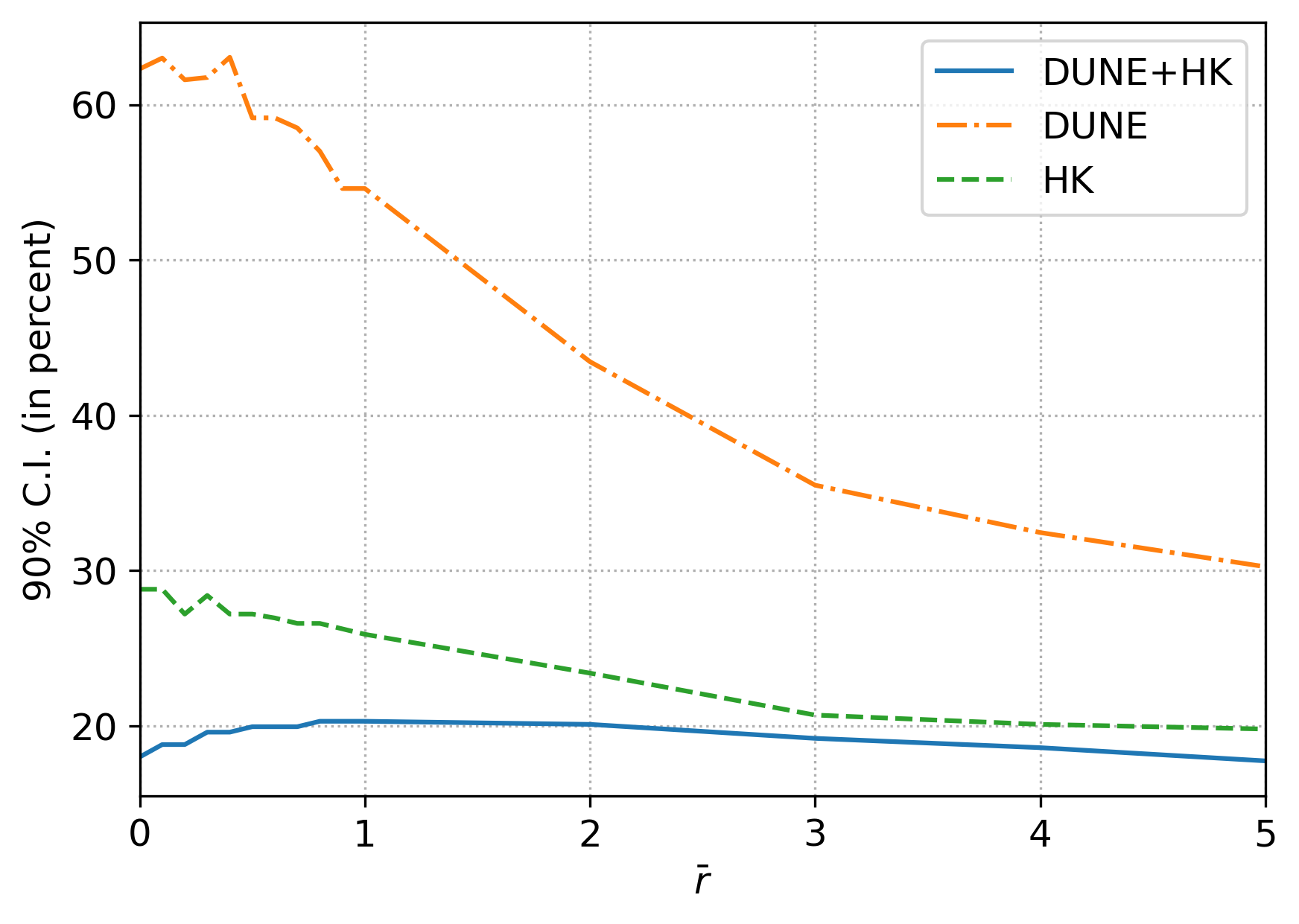}
    \caption{Size of the $90\%$ confidence interval on the measured CCSN distance as a function of $\bar{r}$ for $\zeta=1$, normalized by the true distance, for a supernova at 10~kpc with a $11~M_\odot$ progenitor. The mass ordering is assumed to be the NMO. The experiments shown here are DUNE (dash-dotted orange), Hyper-Kamiokande (dashed green), and the combination of the two detectors (solid blue).}
    \label{fig:CLdistance_rbar_nmo}
\end{figure*}

Figure~\ref{fig:distance_BSM_nmo} shows the median measured CCSN distance and the $90\%$ confidence interval as a function of the true distance for the $(\bar{r}=5,\zeta=1)$ scenario considered in Section~\ref{sec:discussion}, in the NMO, for the DUNE and HK detectors as well as for their combination. In this Figure, the distance estimated using the SM-only likelihood is compared to the distance obtained by allowing $\bar{r}$ and $\zeta$ to vary. 
For all detectors, a bias towards distances smaller than the true value is observed for the SM-only likelihood. However, this bias is milder than for the IMO, with a measured distance of $9$~kpc found for a CCSN located at $10$~kpc. As in Section~\ref{sec:distance}, we then study how the size of the $90\%$ C.I. varies with $\bar{r}$ for DUNE, HK, and DUNE+HK. Again, here, we keep $\zeta$ fixed to one as, for lower values of $\zeta$, the CCSN neutrino flux will be suppressed leading to effects similar to the ones observed in the IMO. The dependence of the CCSN distance measurement precision in $\bar{r}$ is shown in Figure~\ref{fig:CLdistance_rbar_nmo}.  For both HK and DUNE, we now observe a continuous decrease in the size of the uncertainty as $\bar{r}$ grows. As with the IMO, combining data from both DUNE and HK, stabilizes distance measurement uncertainties, maintaining them below $\sim20\%$ and reducing their size by up to $60\%$.

\begin{figure*}
    \centering    \includegraphics[width=\linewidth]{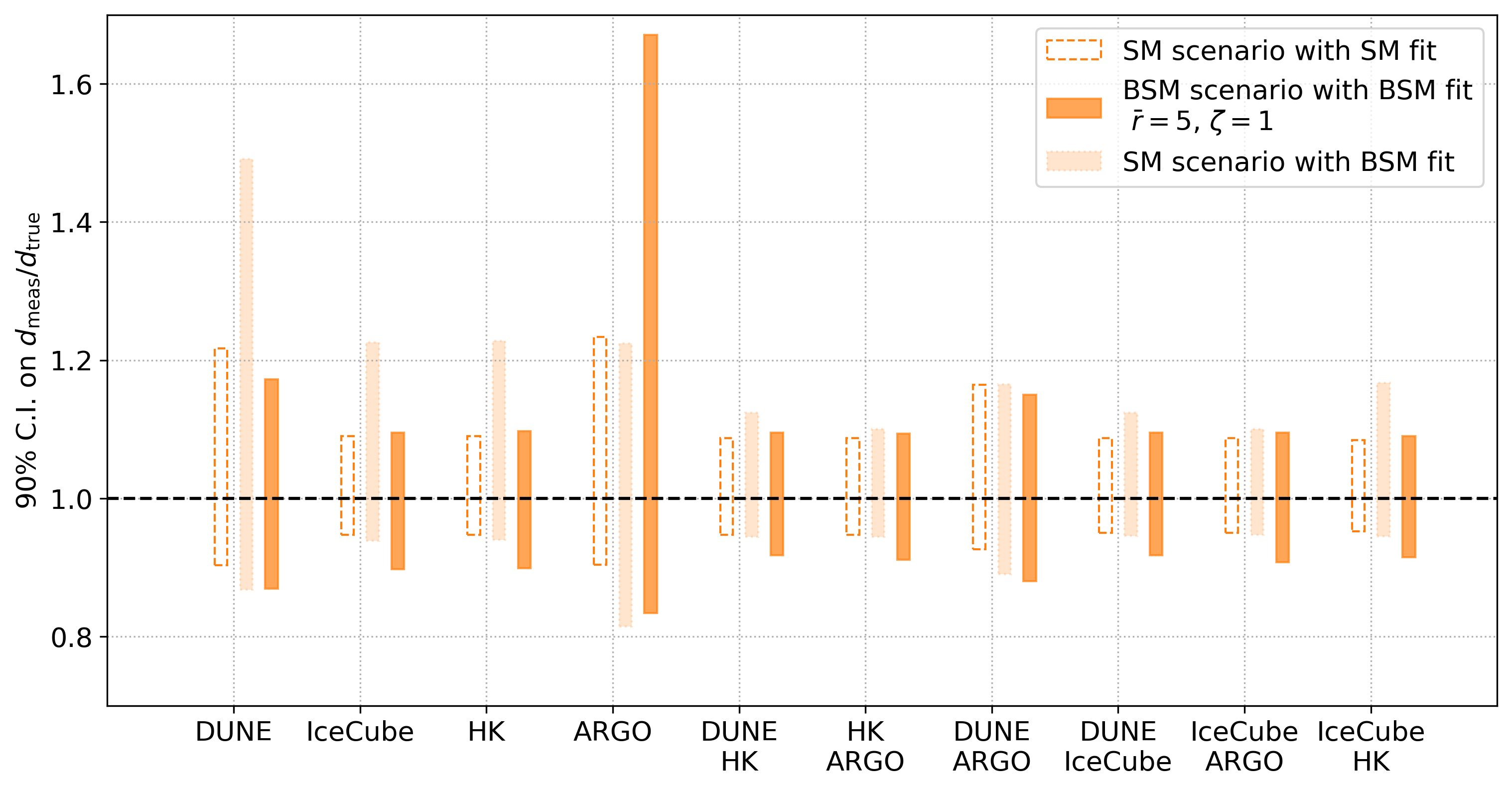}
    \caption{$90\%$ confidence intervals on the measured CCSN distance, normalized by the true distance, for a supernova at 10~kpc, for a $11M_\odot$ progenitor and different detector combinations. The hollow rectangles with a dashed border show the confidence intervals for the SM under the SM hypothesis. The light and dark orange rectangles show the confidence intervals for the SM and for a neutrino decay model with $\bar{r}=5,\zeta=1$, respectively, under the BSM assumption: both measurements are fitted by optimizing the CCSN progenitor mass and location, and the $\bar{r},\zeta$ parameters. The NMO is assumed in both situations. Different detectors and pairs of detectors are considered.}
    \label{fig:CLdistance_BSM_nmo}
\end{figure*}

To investigate variations of the CCSN distance precision with the choice of experiment, we compute the $90\%$ C.I. on the CCSN distance for the individual experiments and detector pairs considered in Figure~\ref{fig:CLdistance_BSM}. In Figure~\ref{fig:CLdistance_BSM_nmo}, we compare these confidence intervals for a SM measurement, both under the SM and BSM assumptions. We also revisit the scenario discussed earlier, $\bar{r}=5$ and $\zeta=1$, under the BSM assumption. For DUNE, HK, and IceCube, we observe again that the uncertainties decrease as $\bar{r}$ increases. Conversely, for ARGO, as in the IMO case, the uncertainties grow as $\bar{r}$ increases. This difference in behaviors can be explained by the fact that, in the NMO, neutrino fluxes increase with $\bar{r}$ at WC detectors and decrease at ARGO, as can be seen in Figure~\ref{fig:decay_snrates}. Here, combining detectors that are sensitive to different neutrino flavors proves effective in mitigating the growth of uncertainties at low $\bar{r}$, with reductions of up to $60\%$ with all combinations for $\bar{r}=0$. For all detector pairs considered, the resulting uncertainty bands are now comparable to the ones expected in the SM with the SM-only likelihood, for both high and low $\bar{r}$. A notable finding for this ordering is the importance of ARGO. While the role of this detector was limited in the IMO, its effect on mitigating measurement uncertainties in the NMO is now similar to the one of DUNE. This improved performance can be explained by the opposite evolution of the neutrino fluxes with $\bar{r}$ at ARGO compared to the other detectors. Indeed, since the neutrino flux normalization is strongly correlated with the CCSN distance, this opposite evolution can significantly narrow down the range of possible CCSN locations.

\bibliographystyle{JHEP}
\bibliography{JHEP}
\end{document}